\documentclass[journal]{IEEEtran}

\usepackage{cite}
\usepackage{graphicx}
\usepackage{amsmath}
\usepackage{amssymb}
\usepackage{subcaption}
\usepackage{color}
\usepackage{array}
\usepackage{mathtools}
\usepackage{algorithm}
\usepackage{algorithmic}
\usepackage{easyReview}
\usepackage{stfloats}

\DeclareUnicodeCharacter{03B1}{\ensuremath{\alpha}} 
\DeclareUnicodeCharacter{03BC}{\ensuremath{\mu}}

\begin{document}
	
	\title{Secure Transmission for THz-Empowered RIS-Assisted Non-Terrestrial Networks}
	
	\author{Jing~Yuan, Gaojie~Chen,~\emph{Senior Member,~IEEE},  Miaowen~Wen,~\emph{Senior Member,~IEEE}, Rahim Tafatzolli,~\emph{Senior Member,~IEEE}, and Erdal~Panayirci,~\emph{Life Fellow,~IEEE}

		\thanks{Jing~Yuan and Miaowen~Wen are with the School of Electronic and Information Engineering, South China University of Technology, Guangzhou 510641, China (e-mail: eeyj@mail.scut.edu.cn; eemwwen@scut.edu.cn).}
		
		\thanks{Gaojie~Chen and  Rahim Tafatzolli are with the 5GIC $\&$ 6GIC, Institute for Communication Systems, University of Surrey, GU2 7XH Guildford, U.K. (e-mail: gaojie.chen@surrey.ac.uk; r.tafazolli@surrey.ac.uk).}
		
		\thanks{Erdal~Panayirci is with the Department of Electrical and Electronics Engineering, Kadir Has University, 34083 Istanbul, Turkey (e-mail: eepanay@khas.edu.tr).}
		
	}

	\maketitle
	
	\begin{abstract}
		The non-terrestrial networks (NTNs) are recognized as a key component to provide cost-effective and high-capacity ubiquitous connectivity in the future wireless communications. In this paper, we investigate the secure transmission in a  terahertz (THz)-empowered reconfigurable intelligent surface (RIS)-assisted NTN (T-RANTN), which is composed of a low-Earth orbit satellite transmitter, an RIS-installed high-altitude platform (HAP) and two unmanned aerial vehicle (UAV) receivers, only one of which is trustworthy. An approximate ergodic secrecy rate (ESR) expression is derived when the atmosphere turbulence and pointing error due to the characteristics of THz as well as the phase errors resulting from finite precision of RIS and imperfect channel estimation are taken into account simultaneously. Furthermore, according to the statistical and perfect channel state information of the untrustworthy receiver, we optimize the phase shifts of RIS to maximize the lower bound of secrecy rate (SR) and instantaneous SR, respectively, by using semidefinite relaxation method. Simulation results show that both the approximate expression for the ESR and the optimization algorithms are serviceable, and even when the jitter standard variance of the trustworthy receiver is greater than that of the untrustworthy one, a positive SR can still be guaranteed.
	\end{abstract}
	

	\IEEEpeerreviewmaketitle

	\section{Introduction}
	
	In recent years, with the global proposal and promotion of integrated space-Earth networks in the beyond fifth-generation (B5G) era, a number of advanced technologies involved, such as unmanned driving and transoceanic communication, have put forward higher requirements for future wireless communication networks, which include uninterrupted and ubiquitous connectivity, as well as ultra-high data rates and reliability \cite{Giordani2021Non}. Nevertheless, it is impractical to achieve these goals by intensively deploying a large number of terrestrial base stations due to their high installation and maintenance costs. 
	
	So far, the non-terrestrial networks (NTNs) may be one of the best solutions to this problem. Specifically, an NTN can consist of terrestrial and non-terrestrial nodes that are within the atmosphere or in the space. For example, a dense low-Earth orbit (LEO) satellite constellation was considered in \cite{Leyva2021Inter}, where an algorithm to dynamically establish the inter-plane inter-satellite links (ISLs) is proposed to maximize the sum rate of the constellation. The authors in \cite{Le2022Throughput} investigated the throughput of a LEO satellite-assisted Internet of vehicles composed of a LEO satellite constellation, multiple transmission control protocol sources and receivers, as well as an unmanned aerial vehicle (UAV) receiver. Therein, the high-data-rate free-space optics (FSO) and various optimization methods for optimal UAV's parameters were used to obtain a maximal system throughput. Furthermore, to serve users in remote areas, the idea of LEO satellites in combination with HAPs by utilizing millimeter wave (mmWave) was proposed in \cite{Jia2021Joint}, where a two-tier matching algorithm is put forward to solve the problem of dynamic connection between HAPs and satellites resulting from the periodic motion of satellites to maximize the revenue in LEO satellites. In \cite{Saeed2021Point,Kaushal2017Optical}, hybrid RF/FSO systems, where the operating frequency could be switched according to the weather and pointing conditions, were studied. Instead of commonly used laser, mmWave and optical waves, the authors in \cite{Kokkoniemi2021Channel} took advantage of the terahertz (THz) band for the wireless connectivity in airplanes, which may be far away from both the satellites and the ground base stations. A detailed channel model for aerial THz communications was proposed and used to prove the feasibility of airborne THz link. It is worth noting that the THz wave supports higher throughput with less beam divergence than mmWave band, and offers a wider beam with an advantage of higher tolerance in pointing error  caused by the sharp beam of high-frequency waves and the relative motion of two communicating equipments than the laser and optical waves counterparts \cite{Ntontin2017Toward,Bhardwaj2021Performance,Andrews1999Theory,Ammar2001Mathematical}.
	
	A line-of-sight (LoS) link is always required to ensure smooth connections. However, the high-frequency waves can be easily blocked by obstructions. The emerging technology,  reconfigurable intelligent surface (RIS), is expected to be one of the most promising  solutions to this issue \cite{Wu2020towards}. From a macro perspective, RIS is a planar array consisting of a large number of elements, which are able to adaptively adjust their amplitudes and phase shifts to change the intensity and direction of the incident waves. It is more attractive that this operation is almost passive compared to traditional relays. Consequently, RIS-assisted NTNs (RANTNs) have been widely studied \cite{Ye2021Non,Jia2020Ergodic,Tekbiyik2020Reconfigurable}. Specifically, in \cite{Jia2020Ergodic}, an RANTN system was considered, where the source transmits the signal via RIS mounted on an UAV to the destination that cannot be directly reached, and the ergodic capacity of the system was derived. Moreover, the model of RIS assisting the communication of two LEO satellites, which may be on the same or different planes, was presented in \cite{Tekbiyik2020Reconfigurable}. Three types of RIS combination: an individual, multiple independent and multiple consecutive RIS, were considered, and their bit error rates were deduced respectively. In particular, the pointing error was taken into account in \cite{Jia2020Ergodic,Tekbiyik2020Reconfigurable}.
	
	Due to the broadcast nature of wireless communications, the confidential information is very likely to be wiretapped. The traditional solution is to encrypt the confidential message, but this is very inefficient \cite{Li2019Secret}. In recent years, the physical layer security (PLS) appears to be an effective method to improve the secrecy of wireless networks. Furthermore, thanks to the occurrence of RIS, the system PLS has become more flexible. The authors in \cite{Kawai2021QoS} minimized the transmit power under the constraint of the legitimate and wiretap users' quality-of-service (QoS) by jointly optimizing the precoding matrix and phase shifts of RIS. In \cite{Dong2020Enhancing,Wang2022Beamforming,Guan2020Intelligent}, the artificial noise, which is in the null space of the legitimate channel, was made full use of to interrupt the wiretap channel to enhance the system security. However, despite the flexibility brought by RIS, the influence of phase error resulting from the finite precision of RIS or imperfect channel estimation on PLS cannot be neglected \cite{Badiu2020Communication}. The secrecy outage probability and ergodic secrecy rate (ESR) of RIS-assisted end-to-end networks in the presence of phase error were analyzed in \cite{Trigui2021Secrecy,Vega2021Physical,Xu2021Ergodic}, where the phase error is modeled by uniform or Von Mises distribution. Especially, the authors in \cite{Xu2021Ergodic} considered two cases of colluding and non-colluding eavesdroppers.
	
	As the deployment of air communication nodes is pushing forward worldwide, the RIS-assisted PLS can also play a significant role. A system model where the RIS covering the facade of a building bridges the communication between the UAV and the legitimate terrestrial user with an eavesdropper nearby was presented in \cite{Fang2021Joint}. The trajectory and power of UAV and the phase shifts of RIS were jointly optimized to maximize the SR. Following \cite{Fang2021Joint}, the time division multiple access was applied for uplink and downlink communications to enhance the security of the system in \cite{Li2021Robust}. Moreover, the method based on reinforcement learning can also be used to improve the system secrecy \cite{Guo2021Learning}. However, in the existing works, the influence of atmospheric environment on the links in the air was not tackled properly, which may cause wave attenuation and air vehicles' instability. Besides, the phase error was not taken into account, which could be devastating to the system security once it occurs. 
	
	In this paper, we investigate the security issue of a THz-empowered RANTN (T-RANTN), which simultaneously suffers from atmosphere turbulence, pointing error and phase error. The main contributions of this paper are summarized as follows:
	\begin{itemize}
		\item It is the first work to investigate the security performance on T-RANTNs in the presence of atmosphere turbulence and pointing error due to the characteristics of THz as well as phase errors resulting from finite precision of RIS and imperfect channel estimation simultaneously. 
		\item We derive an approximate expression for the ESR and optimize the phase shifts of RIS, given the statistical or perfect channel state information (CSI) of the untrustworthy receiver, to maximize the lower bound of SR and the instantaneous SR by using semidefinite relaxation method, respectively.
		\item Monte Carlo simulations validate the tightness of the approximate expression for ESR and the feasibility of the optimization algorithms. Besides, the relationships between ESR and variable parameters are also investigated, from which we arrive at the most important conclusion, namely all of the ESR, lower bound of SR and instantaneous SR can still be guaranteed to be positive even when the jitter standard variance of the trustworthy receiver is greater than that of the untrustworthy one.
	\end{itemize}
	
	The reminder of this paper is organized as follows.  In Section~II, we introduce the system model, in which the mathematical models of atmosphere turbulence, pointing error and phase error are described. The probability density functions (PDFs) of the received signal-to-noise ratios (SNRs) of two UAV receivers are given in Section~III, and we derive an approximate expression for the ESR. In Section~IV, the optimization algorithms for the phase shifts of RIS to maximize the lower bound of SR and instantaneous SR are proposed, under the assumption that the statistical or perfect CSI of the untrustworthy receiver is available, respectively. The simulation and theoretical results are shown in Section~V, followed by the conclusion in Section~VI.
	
	\emph{Notation}: The lowercase, lowercase bold, and uppercase bold characters $x$, $\mathbf{x}$ and $\mathbf{X}$ represent scalars, vectors, and matrices, respectively. $f_x\left( x \right)$ and $f_{x,y} \left(x,y\right)$ represent the PDF of $x$ and the joint PDF of $x$ and $y$, respectively. $\varGamma\left(x\right)$ and $\varGamma \left(a,x\right)$ are the gamma function and the incomplete upper gamma function, respectively. $K_n\left(x\right)$, $I_0 \left( x \right)$ and $\mathrm{erf}\left(x\right)$ denote the modified Bessel function of the second kind with order $n$, the modified Bessel function of the first kind with order zero and the error function, respectively. $\mathbb{E} \left[x\right]$ and $\mathrm{arg}\left(x\right)$ are the expectation of real number $x$ and the phase of complex number $x$, respectively. $x \sim \mathcal{CN} \left( \mu ,\sigma ^2 \right)$ means that the variable $x$ follows the complex Gaussian distribution with mean $\mu$ and variance $\sigma^2$. $\left| x \right|$ denotes the modulus or absolute value of $x$, and $j$ is the imaginary unit. $\prescript{}{p}F_q\left( a_1,...,a_p;b_1,...,b_q;x \right)$ is the generalized hypergeometric function. $\left[ \mathbf{X} \right] _{i,j}$ and $\left[\mathbf{x}\right]_i$ are the element in the $i$th row and $j$th column of matrix $\mathbf{X}$ and the $i$th element of vector $\mathbf{x}$, respectively. $\mathbb{C}^{m \times n}$ denotes the space of $m \times n$ complex-valued matrices. $\mathrm{tr}(\mathbf{X})$ represents the trace of matrix $\mathbf{X}$.

	\section{System model}	
\begin{figure}[!t]
		\centering
		\includegraphics[width=2.1in]{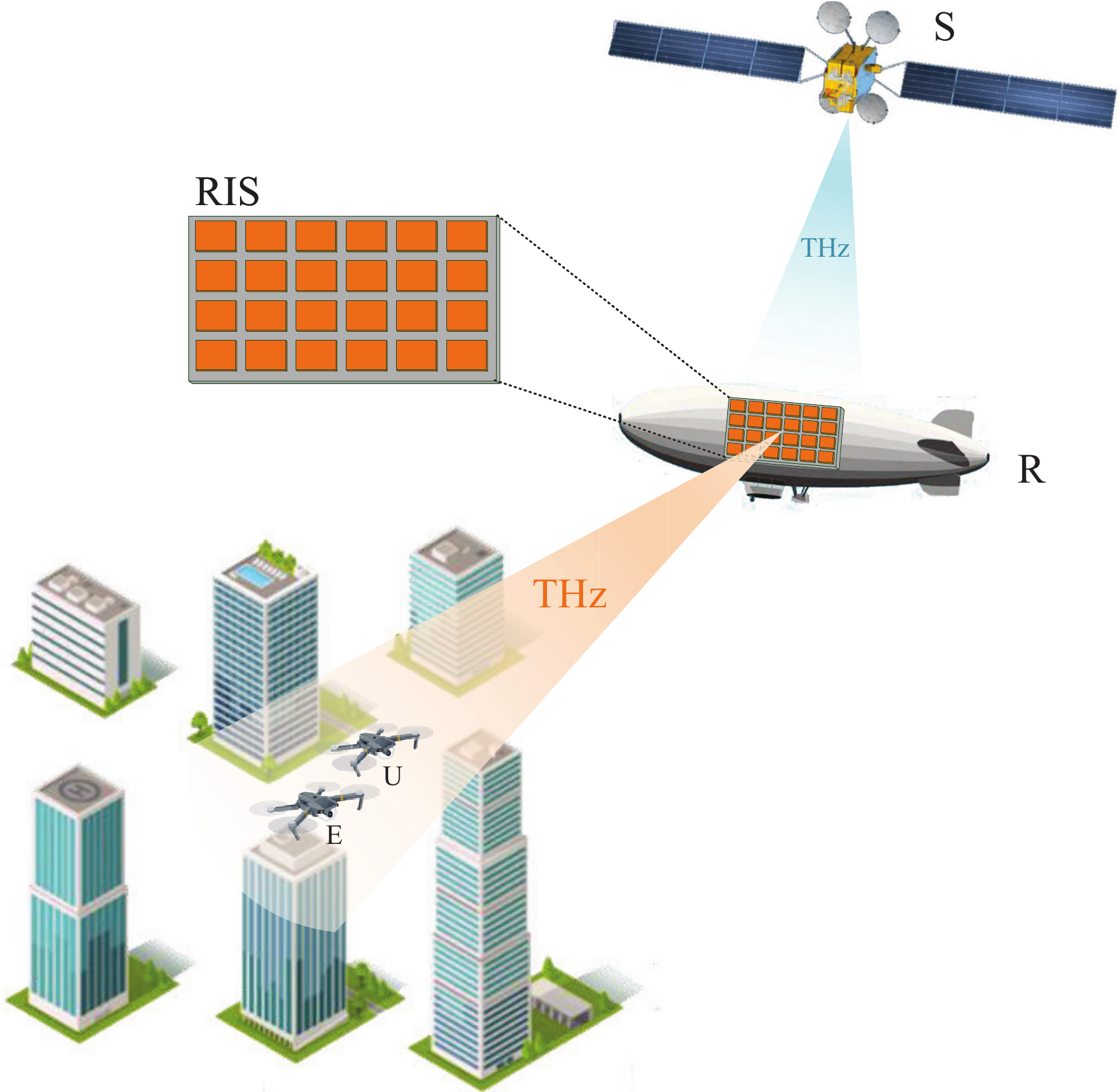}
		\caption{The system model of T-RANTN.}
		\label{fig_system_model}
	\end{figure}

	The system model of T-RANTN is shown in Fig.~\ref{fig_system_model}, where a LEO satellite (S) transmits the signal to a UAV (U) hovering at a low altitude; however, due to the complex surroundings or limited satellite coverage, S is not able to send information to U directly \cite{Li2020A}. Therefore, an HAP (R), mounted with an $N$-element RIS, is utilized to connect S and U. At the same time, there is an untrustworthy UAV (E) located near U, who may want to obtain the confidential information sent from S. Besides, we assume that S, U and E are all equipped with a single antenna because of the hardware limitation \cite{Jia2021Joint}. 
	
	According to the aviation regulations, the HAP is primarily located in the stratosphere, which enables itself to stay at a quasi-stationary position relative to the Earth. Therefore, the periodic movement of the LEO satellite around the Earth will definitely cause a connection from HAP within a certain period of time, and all channel links are assumed to be independent of each other and experience quasi-static fading during this period \cite{Karabulut2021A,Jia2021Joint}. Moreover, due to the severe path loss and high attenuation, the signals reflected by the RIS more than once have negligible power and hence can be ignored \cite{Wu2020towards}.
	
	To achieve higher date rates, the THz band is made full use of in this system. Without loss of generality, we assume ideal LoS channels between S and R with merely path loss owing to the quasi-vacuum communication environment. Furthermore, since both R-U and R-E links are located high in the air with few obstructions, the free-space path loss model is applied. Referring to \cite[Eq.~(2.7)]{Goldsmith2005Wireless}, the common path loss model for S-R, R-U and R-E links is generalized as
	\begin{align}\label{path_loss}
		\mathcal{L} _{\Pi}=G_{\Pi}\left( \frac{\lambda}{4\pi d_{\Pi}} \right) ^2, \, \Pi \in \left\{ s,u,e \right\},
	\end{align}
	where $\lambda$ refers to the wavelength, $G_{\Pi}$ denotes the gain of antennas on S, U and E, and $d_{\Pi}, \, \Pi \in \left\{ s,u,e \right\}$ represents the distance between S and R, R and U and R and E, respectively. 	
	
	In contrast to the aforementioned atmosphere-free circumstance, the UAVs operate at low altitudes of a few hundred meters, where there is troposphere, so that the influence of atmosphere turbulence on THz cannot be ignored \cite{Andrews1999Theory,Ammar2001Mathematical}. In this case, we assume a general atmosphere turbulence fading modeled by Gamma-Gamma distribution as \cite{Jia2020Ergodic,Le2021Performance}
	\begin{align} \label{f_a}	
		f_{\mathcal{T} _{\varrho}}\left( \mathcal{T} _{\varrho} \right) &=\frac{2\left( \alpha _{\varrho}\beta _{\varrho} \right) ^{\frac{ \alpha _{\varrho}+\beta _{\varrho}} {2}}}{\varGamma \left( \alpha _{\varrho} \right) \varGamma \left( \beta _{\varrho} \right)}\mathcal{T} _{\varrho}^{\frac{ \alpha _{\varrho}+\beta _{\varrho}}{2}} K_{\alpha _{\varrho}-\beta _{\varrho}}\left( 2\sqrt{\alpha _{\varrho}\beta _{\varrho}\mathcal{T} _{\varrho}} \right) , \nonumber \\
		& \quad \quad \quad \quad \quad \quad \quad \quad \quad \quad \quad  \mathcal{T} _{\varrho}>0,\,\varrho \in \left\{ u,e \right\},
	\end{align}
	where $\mathcal{T} _{\varrho}$ represents the fading due to atmosphere turbulence on the R-U and R-E links, respectively, $\alpha_\varrho$ and $\beta_\varrho$ are the large-scale and small-scale scintillation parameters, which are obtained by
	\begin{align}
		\alpha_\varrho =\left[ \exp \left( \frac{0.49\varsigma _{\varrho} ^2}{\left( 1+1.11 \varsigma _{\varrho} ^{12/5} \right) ^{7/6}} \right) -1 \right] ^{-1}, 
		\nonumber
	\end{align}
	\begin{align}
		\beta_\varrho =\left[ \exp \left( \frac{0.51 \varsigma _{\varrho} ^2}{\left( 1+0.69 \varsigma _{\varrho} ^{12/5} \right) ^{5/6}} \right) -1 \right] ^{-1},
	\end{align}	
	respectively, and $\varsigma _{\varrho} ^2=1.23\left( 2\pi /\lambda \right) ^{7/6}d_{\varrho}^{11/6}C_{n}^{2} $ with $C_n^2$ denoting the index of refraction structure parameter. 
	
	\begin{figure}[!t]
		\centering
		\includegraphics[width=1.5in]{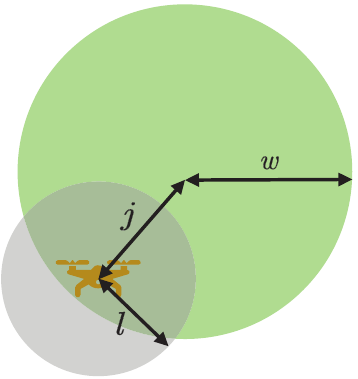}
		\caption{The illustration of pointing error.}
		\label{fig_misalign_fad}
	\end{figure}
	
	Due to air flow and hardware limitation of the device, there is relative motion between S and R, R and U as well as R and E definitely, whereas the sharp beam of THz waves requires a high precision to aim at the effective area of the receiver. Therefore, we must take full account of the pointing error for all links. However, the narrow beam divergence does not pose a threat to the S-R link, because the area of RIS is broad enough for covering the drifts caused by the vibration of R \cite{Jia2020Ergodic,Najafi2021Intelligent}. Without loss of generality, we make a circular beam assumption for the receive antenna's effective area and the transmit beam's footprint. As shown in Fig.~\ref{fig_misalign_fad}, $l$, $w$ and $j$ refer to the radius of the effective area of receive antenna, the beam waist at the receiver and the deviation between the centers of these two areas, respectively. Therefore, referring to \cite{Tekbiyik2020Reconfigurable}, the PDF of the pointing error is expressed as
	\begin{align} \label{f_p}
		f_{\mathcal{P} _{\varrho}}\left( \mathcal{P} _{\varrho} \right) =\frac{\varpi _{\varrho}^{2}}{A_{\varrho}^{\varpi _{\varrho}^{2}}}\mathcal{P} _{\varrho}^{\varpi _{\varrho}^{2}-1},\,0\leqslant \mathcal{P} _{\varrho}\leqslant A_{\varrho}, \, \varrho \in \left\{ u,e \right\},
	\end{align}
	where $\mathcal{P} _{\varrho}$ denotes the fading coefficient brought by the pointing error on R-U and R-E links, $A_{\varrho}$ is the fraction of the collected power at receivers when there is no pointing error
	\begin{align}
		A_{\varrho}=\left[ \mathrm{erf}\left( \frac{\sqrt{\pi}l_{\varrho}}{\sqrt{2}w_{\varrho}} \right) \right] ^2,\nonumber
	\end{align}
	$\varpi_\varrho =W_{\varrho}/2\sigma _{j_\varrho}$ with $\sigma_{j_\varrho}$ denoting the jitter standard variance of U and E, 
	\begin{align}
		W_{\varrho}^{2}=w_{\varrho}^2\frac{\mathrm{erf}\left( \frac{\sqrt{\pi}l_{\varrho}}{\sqrt{2}w_{\varrho}} \right)}{\frac{\sqrt{2}l_{\varrho}}{w_{\varrho}}\exp \left( -\frac{\pi l_{\varrho}^{2}}{2w_{\varrho}^2} \right)} \nonumber
	\end{align}
	is the equivalent beam width at the receivers, $l_\varrho=\lambda \sqrt{G_\varrho}/ \left(2\pi\right)$ and $w_{\varrho}$ are the radius of the effective area of receive antennas and the beam waist on U and E, respectively, $j_\varrho$ is the jitter of U and E.
	
	Moreover, it is intuitive to ignore the small-scale fading when U and E are remote area users as in \cite{Jia2021Joint}. However, even when they are located in complex surroundings, considering the features in THz communications that the power of scattering component is generally much lower (more than 20~dB) than that of LoS component, we can also ignore the small-scale fading and the channels between R and U as well as R and E should be dominated by the LoS component \cite{Ning2021Terahertz,Priebe2013Ultra,Wang2021Outage}. Let $h^r_n=e^{j \theta_n^r}$, $h^u_n=e^{j \theta_n^u}$ and $h^e_n=e^{j \theta_n^e}$, $n \in \left\{ 1,2,...,N \right\}$ denote the normalized LoS channel coefficients of S-R, R-U and R-E links associated with the $n$th element, respectively. Then, with the parameters introduced above, the received signals at U and E can be expressed as follows:
	\begin{align}
		y_u&=\sqrt{P}\left( \sum_{n=1}^N{\sqrt{\mathcal{L} _s}h_{n}^{r} g_n e^{j\upsilon _n}\sqrt{\mathcal{L} _u}\mathcal{T} _u\mathcal{P} _uh_{n}^{u}} \right) x+n_u,
		\label{rec_sig_u} \\
		y_e&=\sqrt{P}\left( \sum_{n=1}^N{\sqrt{\mathcal{L} _s}h_{n}^{r} g_n e^{j\upsilon_n}\sqrt{\mathcal{L} _e}\mathcal{T} _e\mathcal{P} _eh_{n}^{e}} \right)x+n_e, \label{rec_sig_e}
	\end{align}
	where $P$ and $x$ denote the transmit power and the transmitted signal with $\mathbb{E}[ \left| x \right| ^2 ] =1$, respectively, $n_u\sim \mathcal{CN} \left( 0,\delta _{u}^{2} \right)$ and $n_e\sim \mathcal{CN} \left( 0,\delta _{e}^{2} \right)$ are the additive white Gaussian noise (AWGN) with power $\delta _{u}^{2}$ at U and $\delta _{e}^{2}$ at E, respectively, and $g_n$ and $\upsilon_n$ are the amplitude and phase shift of the $n$th element, respectively. In particular, the amplitudes of all elements $g_n, n \in \left\{1,2,...,N\right\}$ are defined as $1$ to ensure the maximal reflect power \cite{Vega2021Physical}. Furthermore, we assume all elements can be adjusted independently.
	
	Ideally, to maximize the received signal power at the target receiver U, the phase shifts of RIS ought to direct all signals toward U to the same direction $\vartheta$. That is, the optimal phase shift of the $n$th element satisfies $\upsilon _n^*= \vartheta -\left( \theta _{n}^{r}+\theta _{n}^{u} \right)$ without requisition of the CSI of S-R-E link. However, in a practical system, although the CSI of S-R-U link is perfectly known, the phase error may also be introduced by the finite adjustment precision of RIS, which is called quantization error. We assume that only a discrete set of $2^b$ phases can be configured where $b \in \left\{1,2,...,N\right\}$ stands for the quantization bit. Mathematically, the resulting phase shift of the $n$th reflecting element is most likely equal to
	\begin{align}\label{equ_phase_error}
		\upsilon _n= \vartheta -\left( \theta _{n}^{r}+\theta _{n}^{u} \right) +\varepsilon _n,
	\end{align}
	where $\varepsilon_n$, $n \in \left\{ 1,2,...,N \right\}$ is the phase error, which is assumed to be uniformly distributed with the PDF as
	\begin{align} \label{PDF_uni}
		f_{\varepsilon _n}\left( \varepsilon _n \right) =\begin{cases}
			\frac{2^{b-1}}{\pi},&		\varepsilon _n\in \left[ -\frac{\pi}{2^b},\frac{\pi}{2^b} \right)\\
			0,&		\mathrm{otherwise}\\
		\end{cases}.
	\end{align}
	Therefore, the SNRs at U and E are attained as
	\begin{align}
		\gamma _u&=\gamma _{u}^{0}\left| \sum_{n=1}^N{\mathcal{T} _u\mathcal{P} _ue^{j\varepsilon _n}} \right|^2
		\label{SNR_u},
		\\
		\gamma _e&=\gamma _{e}^{0}\left| \sum_{n=1}^N{\mathcal{T} _e\mathcal{P} _ee^{j\left( \theta _{n}^{e}-\theta _{n}^{u}+\varepsilon _n \right)}} \right|^2
		\label{SNR_e},
	\end{align}
	respectively, where $\gamma _{u}^{0}=P\mathcal{L} _s\mathcal{L} _u/\delta _{u}^{2}$ and $\gamma _{e}^{0}=P\mathcal{L} _s\mathcal{L} _e/\delta _{e}^{2}$.
	
	\section{Performance analysis}
	Both $\gamma_u$ and $\gamma_{e}$ in (\ref{SNR_u}) and (\ref{SNR_e}) can be disassembled into three parts: the constant terms $\gamma_u^0 N^2$ and $\gamma_e^0$, the combined effect of atmosphere turbulence and pointing error denoted by $h_\varrho=\mathcal{T}_\varrho \mathcal{P}_\varrho, \, \varrho \in \left\{u,e\right\}$, and the equivalent channels of S-R-U and S-R-E links which are represented as $r_u=\frac{1}{N}\left| \sum_{n=1}^N{e^{j\varepsilon _n}} \right|$ and $r_e=\left|\sum_{n=1}^N{e^{j\left( \theta _{n}^{e}-\theta _{n}^{u}+\varepsilon _n \right)}}\right|$, respectively. Therefore, as long as the distribution of $h_\varrho r_\varrho, \, \varrho \in \left\{u,e\right\}$ is derived, those of the received SNRs can be obtained. In the sequel, we will discuss them separately.
	
	\subsection{Distribution of $h_\varrho$}	
	The distribution of $h_\varrho, \, \varrho \in \left\{u,e\right\}$ has been deduced in \cite{Sandalidis2016A} by making use of the approximation of mixture Gamma turbulence model as
	\begin{align} \label{PDF_ht}
		f_{h_{\varrho}}\left( h_{\varrho} \right) &=\frac{\varpi _{\varrho}^{2}}{A_{\varrho}^{\varpi_\varrho ^2}}h_{\varrho}^{\varpi_\varrho ^2-1} \sum_{i=1}^M{a_{\varrho ,i}\xi _{\varrho ,i}^{\varpi _{\varrho}^{2}-\alpha _{\varrho}}\varGamma \left( \alpha _{\varrho}-\varpi _{\varrho}^{2},\frac{\xi _{\varrho ,i}}{A_{\varrho}}h_{\varrho} \right)}, \nonumber \\
		& \quad \quad \quad \quad \quad \quad \quad \quad \quad \quad h_{\varrho}>0, \, \varrho \in \left\{ u,e \right\},
	\end{align}
	where $M$ is the order of the Gauss-Laguerre quadrature method, $a_{\varrho ,i}=\frac{\eta _{\varrho ,i}}{\sum\nolimits_{j=1}^M{\eta _{\varrho ,j}\varGamma \left( \alpha _{\varrho} \right) \xi _{\varrho ,j}^{-\alpha _{\varrho}}}}$, $\xi _{\varrho ,i}=\frac{\alpha _{\varrho}\beta _{\varrho}}{c_{\varrho ,i}}$, $\eta _{\varrho ,i}=\frac{\left( \alpha _{\varrho}\beta _{\varrho} \right) ^{\alpha _{\varrho}}g_{\varrho ,i}c_{\varrho ,i}^{-\alpha _{\varrho}+\beta _{\varrho}-1}}{\varGamma \left( \alpha _{\varrho} \right) \varGamma \left( \beta _{\varrho} \right)}$, and $c_{\varrho,i}$ and $g_{\varrho,i}$ are the abscissas and weight factors for Laguerre integration \cite[Table 25.9]{Milton1972Handbook}.

	\subsection{Distribution of $r_u$}
	Let us rewrite $r_u$ as
	\begin{align}
		r_u=\frac{1}{N}\left| \sum_{n=1}^N{e^{j\varepsilon _n}} \right|=\left| C_u+jS_u \right|,
	\end{align}
	where $C_u=\frac{1}{N}\sum_{n=1}^N{\cos \left( \varepsilon _n \right)}$ and  $S_u=\frac{1}{N}\sum_{n=1}^N{\sin \left( \varepsilon _n \right)}$. The expectations and variances of $C_u$ and $S_u$ can be derived as follows:
	\begin{align}\label{E_V_Cu_Su}
		\mu _{u,c}&=\mathbb{E} \left[ C_u \right] =\mathbb{E} \left[ \cos \left( \varepsilon _n \right) \right] =\varphi _1, \nonumber \\
		\mu _{u,s}&=\mathbb{E} \left[ S_u \right] =0,
		\nonumber \\
		\sigma _{u,c}^{2}&=\frac{1}{N}\left( \mathbb{E} \left[ \cos ^2\left( \varepsilon _n \right) \right] -\left( \mathbb{E} \left[ \cos \left( \varepsilon _n \right) \right] \right) ^2 \right) \nonumber \\
		&=\frac{1+\varphi _2-2 \varphi _{1}^{2}}{2N}
		, \nonumber \\
		\sigma _{u,s}^{2}&=\frac{1}{N}\left( \mathbb{E} \left[ \sin ^2\left( \varepsilon _n \right) \right] -\left( \mathbb{E} \left[ \sin \left( \varepsilon _n \right) \right] \right) ^2 \right)
		\nonumber \\
		&=\frac{1-\varphi _2}{2N},
	\end{align} 
	where $\varphi_p, \, p=1,2$ is the characteristic function of $\varepsilon_{n}$,
	\begin{align}
		\varphi _p=\mathbb{E} \left[ e^{jp\varepsilon _n} \right] =\frac{2^b\sin \left( 2^{-b}p\pi \right)}{p\pi}.
	\end{align}
	With a view to the Gaussian distribution followed by $C_u$ and $S_u$ by means of the central limited theorem (CLT) when $N\rightarrow \infty$, the discussion of their independence is equivalent to that of their correlation with the Pearson's correlation coefficient to be
	\begin{align}\label{correlation coefficient}
		\rho =\frac{\mathbb{E} \left[ C_uS_u \right] - \mu_{u,c} \mu_{u,s}}{\sigma _{u,c}\sigma _{u,s}},
	\end{align}
	where 
	\begin{align}\label{E_Cu_Su}
		\mathbb{E} \left[ C_uS_u \right] &=\frac{1}{N^2}\mathbb{E} \left[ \sum_{i=1}^N{\sum_{j=1}^N{\cos \left( \varepsilon _i \right) \sin \left( \varepsilon _j \right)}} \right]
		\nonumber \\
		&=\frac{1}{N^2}\sum_{i=1}^N{\sum_{j=1,j\ne i}^N{\mathbb{E} \left[ \cos \left( \varepsilon _i \right) \sin \left( \varepsilon _j \right) \right]}} \nonumber \\
		& \quad + \frac{1}{N^2}\sum_{i=1}^N{\mathbb{E} \left[ \cos \left( \varepsilon _i \right) \sin \left( \varepsilon _i \right) \right]}.
	\end{align}
	After calculation we find that $\mathbb{E} \left[\cos \left( \varepsilon _i \right) \sin \left( \varepsilon _j \right) \right] = \mathbb{E} \left[ \cos \left( \varepsilon _i \right) \sin \left( \varepsilon _i \right) \right] =0$ always holds. As a result, $\rho = 0$, that is, $C_u$ and $S_u$ are independent. Based on the analysis above, referring to \cite{Badiu2020Communication}, $r_u$ has an approximate Nakagami distribution with PDF as
	\begin{align}\label{PDF_ru}
		f_{r_u}\left( r_u \right) =\frac{2m^m}{\varGamma \left( m \right) \varOmega ^m}r_{u}^{2m-1}e^{-\frac{m}{\varOmega}r_{u}^{2}}, \, r_u\geqslant 0,
	\end{align}
	for large $N$, where $m=\mu _{u,c}^{2}/4\sigma _{u,c}^{2}$ and $\varOmega =\mu _{u,c}^{2}$.

	\subsection{Distribution of $r_e$}	
	Let $\theta _{n}^{sre}=\theta _{n}^{e}-\theta _{n}^{u}+\varepsilon _n$. Then, $r_e$ can be expressed as $r_e=\left|\sum_{n=1}^N{e^{j\theta_n^{sre}}}\right|$. To obtain the distribution of $r_e$, we need to derive the distribution of $\theta_{n}^{sre}$ first. As is generally assumed in the existing works, $\theta_{n}^{e}$ and $\theta_{n}^{u}$ are uniformly distributed in $\left[ 0,2\pi \right) $; therefore, $\theta_{n}^{e} - \theta_{n}^{u}$ is also uniformly distributed in the same interval. Therefore, the PDF of $\theta_n^{sre}$ is equal to the convolution of those of $\theta_n^{e}-\theta_n^{u}$ and $\varepsilon _n$:
	\begin{align} \label{conv_PDF}
		f_{\theta _{n}^{sre}}\left( \theta _{n}^{sre} \right) =\int_{-\infty}^{+\infty}{f_{\theta _{n}^{e}-\theta _{n}^{u}}\left( \tau \right) f_{\varepsilon _n}\left( \theta _{n}^{sre}-\tau \right) \mathrm{d}\tau}.
	\end{align}
	Substituting $f_{\theta _{n}^{e}-\theta _{n}^{u}}\left( \tau \right) =\begin{cases}
		1/2\pi ,&		\tau \in \left[ 0,2\pi \right)\\
		0,&		\mathrm{otherwise}\\
	\end{cases}$ and (\ref{PDF_uni}) into (\ref{conv_PDF}) yields
	\begin{align}\label{PDF_theta_sre1}
		&f_{\theta _{n}^{sre}}\left( \theta _{n}^{sre} \right) =
		\nonumber \\
		&\begin{cases}
			\frac{2^{b-2}}{\pi ^2}\theta _{n}^{sre}+\frac{1}{4\pi},&		\theta _{n}^{sre}\in \left[ -\frac{\pi}{2^b},\frac{\pi}{2^b} \right)\\
			\frac{1}{2\pi},&		\theta _{n}^{sre}\in \left[ \frac{\pi}{2^b},2\pi -\frac{\pi}{2^b} \right)\\
			-\frac{2^{b-2}}{\pi ^2}\theta _{n}^{sre}+\frac{2^{b+1}+1}{4\pi},&		\theta _{n}^{sre}\in \left[ 2\pi -\frac{\pi}{2^b},2\pi +\frac{\pi}{2^b} \right)\\
			0,&		\mathrm{otherwise}\\
		\end{cases}.
	\end{align}
	
	Likewise, we re-define $r_e=\left| C_e+jS_e \right|$, where $C_e = \sum_{n=1}^N{\cos \left( \theta _{n}^{sre} \right)}$ and $S_e = \sum_{n=1}^N{\sin \left( \theta _{n}^{sre} \right)}$. The expectations and variances of $C_e$ and $S_e$ can be calculated as follows:
	\begin{align} \label{E_V_Ce_Se}
		\mu _{e,c}&=\mathbb{E} \left[ C_e \right] =0,
		\nonumber \\
		\mu _{e,s}&=\mathbb{E} \left[ S_e \right] =0,
		\nonumber \\
		\sigma _{e,c}^{2}&=N\left( \mathbb{E} \left[ \cos ^2\left( \theta _{n}^{sre} \right) \right] -\left( \mathbb{E} \left[ \cos \left( \theta _{n}^{sre} \right) \right] \right) ^2 \right) =\frac{N}{2},
		\nonumber \\
		\sigma _{e,s}^{2}&=N\left( \mathbb{E} \left[ \sin ^2\left( \theta _{n}^{sre} \right) \right] -\left( \mathbb{E} \left[ \sin \left( \theta _{n}^{sre} \right) \right] \right) ^2 \right)=\frac{N}{2}.
	\end{align}
	Obviously, $C_e$ and $S_e$ follow the same zero-mean Gaussian distribution for a large value of $N$ in terms of CLT. Moreover, after a similar but more complex process to (\ref{correlation coefficient}) and (\ref{E_Cu_Su}),
	we infer that $C_e$ and $S_e$ are also independent. Therefore, substituting the parameters in (\ref{E_V_Ce_Se}) into the PDF of the two-dimensional Gaussian distribution generates
	\begin{align}\label{PDF_Ce_Se}
		f_{C_e,S_e}\left( x,y \right) =\frac{1}{2\pi \sigma _{e,c}\sigma _{e,s}}e^{-\frac{x^2}{2\sigma _{e,c}^{2}}-\frac{y^2}{2\sigma _{e,s}^{2}}}.
	\end{align}
	For the PDF of $r_e$, we can apply the transformation $f_{r_e,\phi}\left( r_e,\phi \right) =r_ef_{C_e,S_e}\left( x,y \right)$ to (\ref{PDF_Ce_Se}), where $\phi$ is the phase of $\sum_{n=1}^N{e^{j\theta _{n}^{sre}}}$ satisfying $C_e=r_e\cos \left( \phi \right)$ and $S_e=r_e\sin \left( \phi \right)$. Therefore, (\ref{PDF_Ce_Se}) can be transformed into
	\begin{align}
		f_{r_e,\phi}\left( r_e,\phi \right) =\frac{r_e}{2\pi \sigma _{e,c}\sigma _{e,s}}e^{-\frac{ r_e^2\cos^2 \left( \phi \right)}{2\sigma _{e,c}^{2}}-\frac{ r_e^2\sin^2 \left( \phi \right)}{2\sigma _{e,s}^{2}}}.
	\end{align}
	One of the marginal distributions is the PDF of $r_e$, which can be obtained as
	\begin{align}\label{PDF_re}
		f_{r_e}\left( r_e \right) =\frac{r_e}{\sigma _{e,c}^{2}}e^{-\frac{r_{e}^{2}}{2\sigma _{e,c}^{2}}}, \, r_e\geqslant 0.
	\end{align}
	
	\subsection{Ergodic Secrecy Rate}
	The last step before calculating the ESR is to derive the distributions of $H_u = h_u r_u$ and $H_e = h_e r_e$. The PDF of $H_u$ can be readily derived as
	\begin{align}\label{f_Hu}
		f_{H_u}\left( z \right) &=\int_0^{\infty}{\frac{f_{h_u}\left( x \right)}{x}f_{r_u}\left( \frac{z}{x} \right) \mathrm{d}x}
		\nonumber \\
		& =\sum_{i=1}^M{\frac{\varpi _{u}^{2}}{A_{u}^{\varpi_u ^2}}\frac{2m^m}{\varGamma \left( m \right) \varOmega ^m}a_{u,i}\xi _{u,i}^{\varpi _{u}^{2}-\alpha _u}I_1},
	\end{align}
	where $I_1$ is shown in (\ref{I1_clo}) on the top of the next page. 
	\begin{figure*}[!t] 
		\begin{align} \label{I1_clo}
			I_1 &= z^{2m-1}\left( \mathcal{I}_1^1 + \mathcal{I}_1^2 - \mathcal{I}_1^3 + \mathcal{I}_1^4 \right), \nonumber \\
			\mathcal{I} _1^1&=\frac{1}{2}\left( \frac{\varOmega}{mz^2} \right) ^{m-\frac{\varpi _{u}^{2}}{2}}\varGamma \left( \alpha _u-\varpi _{u}^{2} \right) \varGamma \left( m-\frac{\varpi _{u}^{2}}{2} \right), \nonumber \\
			\mathcal{I} _1^2&=\frac{1}{\varpi _{u}^{2}-2m}\left( \frac{\xi _{u,i}}{A_u} \right) ^{2m-\varpi _{u}^{2}}\varGamma \left( \alpha _u-2m \right) \prescript{}{1}F_3\left( m-\frac{\varpi _{u}^{2}}{2};\frac{1-\alpha _u}{2}+m,1-\frac{\alpha _u}{2}+m,1+m-\frac{\varpi _{u}^{2}}{2};-\frac{\xi _{u,i}^{2}mz^2}{4A_{u}^{2}\varOmega} \right),  \nonumber\\
			\mathcal{I} _1^3&=\frac{\left( \frac{\xi _{u,i}}{A_u} \right) ^{\alpha _u-\varpi _{u}^{2}}\left( \frac{\varOmega}{mz^2} \right) ^{-\frac{\alpha _u}{2}+m}}{2\left( \alpha _u-\varpi _{u}^{2} \right)}\varGamma \left( m-\frac{\alpha _u}{2} \right) \prescript{}{1}F_3\left( \frac{\alpha _u-\varpi _{u}^{2}}{2};\frac{1}{2},\frac{2+\alpha _u}{2}-m,\frac{2+\alpha _u-\varpi _{u}^{2}}{2};-\frac{\xi _{u,i}^{2}mz^2}{4A_{u}^{2}\varOmega} \right),  \nonumber\\
			\mathcal{I} _{1}^{4}&=\frac{\hspace{-0.1cm}\left( \frac{\xi _{u,i}}{A_u} \right) ^{1+\alpha _u\hspace{-0.05cm}-\varpi _{u}^{2}} \hspace{-0.1cm} \left( \frac{\varOmega}{mz^2} \right) ^{\hspace{-0.1cm}-\frac{1+\alpha _u}{2}+m}}{2\left( 1+\alpha _u-\varpi _{u}^{2} \right)} \varGamma \hspace{-0.1cm} \left( m\hspace{-0.1cm}-\hspace{-0.1cm}\frac{1+\alpha _u}{2} \right) \prescript{}{1}F_3\left( \frac{1+\alpha _u-\varpi _{u}^{2}}{2};\frac{3}{2},\frac{3+\alpha _u}{2}\hspace{-0.1cm}-\hspace{-0.1cm}m,\frac{3+\alpha _u-\varpi _{u}^{2}}{2};-\frac{\xi _{u,i}^{2}mz^2}{4A_{u}^{2}\varOmega} \right).
		\end{align}
	\hrulefill 
	\end{figure*} 
	Similarly, the PDF of $H_e$ can also be attained as
	\begin{align}\label{f_He}
		f_{H_e}\left( z \right) =\sum_{i=1}^M{\frac{\varpi _{e}^{2}}{A_{e}^{\varpi _{e}^{2}}\sigma _{e,c}^{2}}a_{e,i}\xi _{e,i}^{\varpi _{e}^{2}-\alpha _e}I_2},
	\end{align}
	where $I_2$ is shown in (\ref{I2_clo}) on the top of the next page.
	\begin{figure*}[!t]
		\begin{align} \label{I2_clo}
			I_2&=z\left( \mathcal{I}_{2}^{1}+\mathcal{I}_{2}^{2}-\mathcal{I}_{2}^{3}+\mathcal{I}_{2}^{4} \right),
			\nonumber \\
			\mathcal{I} _{2}^{1}&=2^{-\frac{\varpi _{e}^{2}}{2}}\left( \frac{\sigma _{e,c}^{2}}{z^2} \right) ^{1-\frac{\varpi _{e}^{2}}{2}}\varGamma \left( \alpha _e-\varpi _{e}^{2} \right) \varGamma \left( 1-\frac{\varpi _{e}^{2}}{2} \right), 
			\nonumber \\
			\mathcal{I} _{2}^{2}&=\frac{1}{-2+\varpi _{e}^{2}}\left( \frac{\xi _{e,i}}{A_e} \right) ^{2-\varpi _{e}^{2}}\varGamma \left( -2+\alpha _e \right) \prescript{}{1}F_3\left( 1-\frac{\varpi _{e}^{2}}{2};\frac{3}{2}-\frac{\alpha _e}{2},2-\frac{\alpha _e}{2},2-\frac{\varpi _{e}^{2}}{2};-\frac{\xi _{e,i}^{2}z^2}{8A_{e}^{2}\sigma _{e,c}^{2}} \right),
			\nonumber \\
			\mathcal{I} _{2}^{3}&=\frac{2^{-\frac{\alpha _e}{2}}}{\alpha _e-\varpi _{e}^{2}}\left( \frac{\xi _{e,i}}{A_e} \right) ^{\alpha _e-\varpi _{e}^{2}}\left( \frac{\sigma _{e,c}^{2}}{z^2} \right) ^{1-\frac{\alpha _e}{2}}\varGamma \left( 1-\frac{\alpha _e}{2} \right) \prescript{}{1}F_3\left( \frac{\alpha _u}{2}-\frac{\varpi _{e}^{2}}{2};\frac{1}{2},\frac{\alpha _e}{2},1+\frac{\alpha _u}{2}-\frac{\varpi _{e}^{2}}{2};-\frac{\xi _{e,i}^{2}z^2}{8A_{e}^{2}\sigma _{e,c}^{2}} \right) ,
			\nonumber \\
			\mathcal{I} _{2}^{4}&=\frac{2^{-\frac{1+\alpha _e}{2}}}{1+\alpha _e-\varpi _{e}^{2}} \hspace{-0.1cm} \left( \frac{\xi _{e,i}}{A_e} \right) ^{1+\alpha _e-\varpi _{e}^{2}} \hspace{-0.1cm} \left( \frac{\sigma _{e,c}^{2}}{z^2} \right) ^{\frac{1-\alpha _e}{2}}\hspace{-0.3cm}\varGamma \hspace{-0.1cm} \left( \frac{1-\alpha _e}{2} \right) \prescript{}{1}F_3 \hspace{-0.1cm} \left( \frac{1+\alpha _e \hspace{-0.05cm} - \hspace{-0.05cm} \varpi _{e}^{2}}{2};\frac{3}{2},\frac{1+\alpha _e}{2},\frac{3+\alpha _e-\varpi _{e}^{2}}{2};-\frac{\xi _{e,i}^{2}z^2}{8A_{e}^{2}\sigma _{e,c}^{2}} \right).
		\end{align}
	\hrulefill 
	\end{figure*}
	
	As stated before, the received SNRs of U and E in (\ref{SNR_u}) and (\ref{SNR_e}) can be re-expressed as $\gamma _u=\gamma _{u}^{0}N^2\left(h_ur_u\right)^2=\gamma _{u}^{0}N^2H_u^2$ and $\gamma _e=\gamma _{e}^{0}\left(h_er_e\right)^2=\gamma _{e}^{0}H_e^2$. Therefore, the ESR is generalized as
	\begin{align} \label{ESR_int1}
		R_s=\left[ \mathbb{E} \left[ R_u \right] -\mathbb{E} \left[ R_e \right] \right] ^+,
	\end{align}	
	where 
	\begin{align}
		\mathbb{E} \left[ R_u \right] &=\int_0^{\infty}{\log _2\left( 1+\gamma _{u}^{0}N^2z^2 \right) f_{H_u}\left( z \right) \mathrm{d}z},
		\nonumber \\
		\mathbb{E} \left[ R_e \right] &=\int_0^{\infty}{\log _2\left( 1+\gamma _{e}^{0}z^2 \right) f_{H_e}\left( z \right) \mathrm{d}z}\label{ergodic_rate_1},		
	\end{align}
	denote the ergodic rates of U and E, which can be easily solved by mathematical softwares such as MATLAB through numerical integration. 
	
	\subsection{Considering Both Quantization and Estimation Errors}
	According to \cite{Trigui2021Secrecy,Vega2021Physical,Xu2021Ergodic},  an imperfect channel estimation should also be considered in practical scenarios, and the phase error it brings is called the estimation error. Different from the quantization error attached to $\upsilon_n^*$ mentioned in (\ref{equ_phase_error}), the estimation error comes with $\vartheta$, $\theta _{n}^{r}$ and $\theta _{n}^{u}$ during an imperfect channel estimation. In this subsection, we will analyze the effect of the superposition of these two kinds of phase errors on the ESR.
	
	When the phase error, $\nu_n$, is introduced in the imperfect channel estimation, it follows the Von Mises distribution with zero mean and concentration parameter $\kappa \in \left\{1,2,...,N\right\}$, whose PDF is expressed as \cite{Wang2021Outage}
	\begin{align}\label{PDF_Von_Mises}
		f_{\nu _n}\left( \nu _n \right) =\frac{e^{\kappa \cos \left( \nu _n \right)}}{2\pi I_0\left( \kappa \right)}, ~ \nu _n \in \left[ -\pi,\pi\right).
	\end{align}
	Simultaneously taking quantization and estimation errors into account, the total phase error $\hat{\varepsilon}_n$ is the sum of two independent variables subject to uniform and Von Mises distributions, respectively, namely $\hat{\varepsilon}_n = \varepsilon_{n} + \nu_{n}$. Therefore, $\hat{\varepsilon}_n$ can be analogous to $\varepsilon_n$ in (\ref{equ_phase_error}). The analysis of the theoretical ESR is the same as that of (\ref{ESR_int1}), except for a few differences described below.
	
	Firstly, the distribution of $\hat{\varepsilon}_n$ needs to be derived. Since the PDF of Von Mises distribution (\ref{PDF_Von_Mises}) is not integrable at the endpoints, we first need to make a transformation to it. Considering the case where $\kappa \gg 1$, we have $I_0 \left( \kappa \right) \simeq e^{\kappa}/{\sqrt{2\pi \kappa}}$. Meanwhile, when $\nu_n \rightarrow 0$, which is equivalent to $\kappa \gg 1$, $\cos \left( \nu _n \right) \simeq 1-\nu _{n}^{2}/2$ holds \cite{Gappmair2017Subcarrier}. Substituting  these two equations into (\ref{PDF_Von_Mises}), it is observed that when $\kappa \gg 1$ or $\nu_{n} \rightarrow 0$, the distribution of $\nu_n$ can be re-expressed as
	\begin{align}
		f_{\nu _n}\left( \nu _n \right) \simeq \sqrt{\frac{\kappa}{2\pi}}\exp \left( -\frac{\kappa \nu _{n}^{2}}{2} \right) , \, \nu _n\in \left[ -\pi ,\pi \right).
	\end{align}
	Convolving the PDFs of $\varepsilon_{n}$ and $\nu_{n}$ yields the PDF of $\hat{\varepsilon}_n$, which is shown in (\ref{PDF_new_phase_error}) on the top of the next page.
	\begin{figure*}[!t]
		\begin{align} \label{PDF_new_phase_error}
			f_{\hat{\varepsilon}_n}\hspace{-0.1cm}\left( \hat{\varepsilon}_n \right) \hspace{-0.1cm}=\hspace{-0.1cm}\begin{cases}
				\frac{2^{b-2}}{\pi}\left[ \mathrm{erf}\left( \sqrt{\frac{\kappa}{2}}\pi \right) \hspace{-0.1cm}+\hspace{-0.1cm}\mathrm{erf}\left( \sqrt{\frac{\kappa}{2}}\left( \frac{\pi}{2^b}+\hat{\varepsilon}_n \right) \right) \right] ,&	\hspace{-0.3cm}	\hat{\varepsilon}_n\in \left[ -\pi -\frac{\pi}{2^b},-\pi +\frac{\pi}{2^b} \right)\\
				\frac{2^{b-2}}{\pi}\left[ -\mathrm{erf}\left( \sqrt{\frac{\kappa}{2}}\left( -\frac{\pi}{2^b}+\hat{\varepsilon}_n \right) \right) \hspace{-0.1cm}+\hspace{-0.1cm}\mathrm{erf}\left( \sqrt{\frac{\kappa}{2}}\left( \frac{\pi}{2^b}+\hat{\varepsilon}_n \right) \right) \right] ,&	\hspace{-0.3cm}	\hat{\varepsilon}_n\in \left[ -\pi +\frac{\pi}{2^b},\pi -\frac{\pi}{2^b} \right)\\
				\frac{2^{b-2}}{\pi}\left[ \mathrm{erf}\left( \sqrt{\frac{\kappa}{2}}\pi \right) \hspace{-0.1cm}-\hspace{-0.1cm}\mathrm{erf}\left( \sqrt{\frac{\kappa}{2}}\left( -\frac{\pi}{2^b}+\hat{\varepsilon}_n \right) \right) \right] ,&	\hspace{-0.3cm}	\hat{\varepsilon}_n\in \left[ \pi -\frac{\pi}{2^b},\pi +\frac{\pi}{2^b} \right)\\
				0,&	 \hspace{-0.3cm}	\mathrm{else}\\
			\end{cases}\hspace{-0.3cm}.
		\end{align}
	\hrulefill
	\end{figure*}
	Furthermore, since the distribution of $\hat{\varepsilon}_n$ is symmetric with respect to the vertical axis, its characteristic function must be in the real number field, that is,
	\begin{align}
		\hat{\varphi} _1&=\mathbb{E} \left[ e^{j\hat{\varepsilon} _n} \right] =\mathbb{E} \left[\mathrm{cos}(\hat{\varepsilon}_{n})\right],
		\nonumber \\
		\hat{\varphi} _2&=\mathbb{E} \left[ e^{j2\hat{\varepsilon} _n} \right] =\mathbb{E} \left[\mathrm{cos}(2\hat{\varepsilon}_{n})\right],
	\end{align}
	which can be calculated by mathematical softwares and substituted into (\ref{E_V_Cu_Su}) to obtain $\hat{\mu}_{u,c}$, $\hat{\mu}_{u,s}$, $\hat{\sigma}_{u,c}^2$ and $\hat{\sigma}_{u,s}^2$.
	
	On the other hand, the phase related to the $n$th element of the S-R-E link in this case can be expressed as $\hat{\theta} _{n}^{sre}=\theta _{n}^{e}-\theta _{n}^{u}+\varepsilon_{n}+\nu _n$. It is observed that the PDF of $\hat{\theta}_{sre}$ can be derived by convolving those of $\theta _{n}^{e}-\theta _{n}^{u}+\varepsilon_{n}$ in (\ref{PDF_theta_sre1}) and $\nu_{n}$ in (\ref{PDF_Von_Mises}). Therefore, we can obtain the PDF of $\hat{\theta}_{sre}$ as
	\begin{align}
		&f_{\hat{\theta} _{n}^{sre}}\left( \hat{\theta} _{n}^{sre} \right) =
		\nonumber \\
		&\begin{cases}
			\frac{\mathrm{erf}\left( \sqrt{\frac{\kappa}{2}}\pi \right) +\mathrm{erf}\left( \sqrt{\frac{\kappa}{2}}\hat{\theta}_{n}^{sre} \right)}{4\pi},&		\hat{\theta}_{n}^{sre}\in \left[ -\pi ,\pi \right)\\
			\frac{\mathrm{erf}\left( \sqrt{\frac{\kappa}{2}}\pi \right) +\mathrm{erf}\left( \sqrt{\frac{\kappa}{2}}\left( 2\pi -\hat{\theta}_{n}^{sre} \right) \right)}{4\pi},&		\hat{\theta}_{n}^{sre}\in \left[ \pi ,3\pi \right)\\
			0,&		\mathrm{otherwise}\\
		\end{cases}.
	\end{align}
	Substitute $\hat{\theta}_{n}^{sre}$ for $\theta_{n}^{sre}$ in  $r_e=\left|\sum_{n=1}^N{e^{j\theta_n^{sre}}}\right|$, and then the parameters $\hat{\mu}_{e,c}$, $\hat{\mu}_{e,s}$, $\hat{\sigma}_{e,c}^2$ and $\hat{\sigma}_{e,s}^2$ corresponding to those in (\ref{E_V_Ce_Se}) are deduced to be
	\begin{align}
		\hat{\mu}_{e,c}&=\hat{\mu}_{e,s}=0, \nonumber \\
		\hat{\sigma}_{e,c}^2&=\hat{\sigma}_{e,s}^2=\frac{N}{2} \mathrm{erf}\left(\sqrt{\frac{\kappa}{2}}\pi\right).
	\end{align}
	
	It should be noted that $C_u$ and $S_u$ as well as $C_e$ and $S_e$ are also independent in this case. To this end, the approximate ESR expression in the presence of both quantization and estimation errors can be also obtained as
	\begin{align}
		\hat{R}_s=\left[ \mathbb{E} \left[ \hat{R}_u \right] -\mathbb{E} \left[ \hat{R}_e \right] \right] ^+,
	\end{align}
	with
	\begin{align}
		\mathbb{E} \left[ \hat{R}_u \right] &=\int_0^{\infty}{\log _2\left( 1+\gamma _{u}^{0}N^2z^2 \right) \hat{f}_{H_u}\left( z \right) \mathrm{d}z},
		\nonumber \\
		\mathbb{E} \left[ \hat{R}_e \right] &=\int_0^{\infty}{\log _2\left( 1+\gamma _{e}^{0}z^2 \right) \hat{f}_{H_e}\left( z \right) \mathrm{d}z},
	\end{align}
	where $\hat{f}_{H_u}\left( z \right)$ is attained by replacing $\mu_{u,c}$ and $\sigma_{u,c}^2$ with $\hat{\mu}_{u,c}$ and $\hat{\sigma}_{u,c}^2$ in (\ref{PDF_ru}) to get $\hat{f}_{r_u}(r_u)$, respectively, and then use $\hat{f}_{r_u}(r_u)$ in (\ref{f_Hu}). Similarly, $\hat{f}_{H_e}\left( z \right)$ is acquired by replacing $\sigma_{e,c}^2$ with $\hat{\sigma}_{e,c}^2$ in (\ref{PDF_re}) to get $\hat{f}_{r_e}(r_e)$, and then use $\hat{f}_{r_e}(r_e)$ in (\ref{f_He}).
	
	\section{Problem Formulation and Optimization}
	In this section, we optimize the phase shifts of RIS to maximize the SR based on the statistical and perfect CSI of E, respectively.
	
	\subsection{Problem Formulation}
	Rewrite (\ref{rec_sig_u}) and (\ref{rec_sig_e}) as 
	\begin{align}
		y_u&=\sqrt{P\mathcal{L} _s\mathcal{L} _u}\mathcal{T} _u\mathcal{P} _u\mathbf{h}_{u}^{H}\mathbf{\Theta h}_r+n_u,
		\\
		y_e&=\sqrt{P\mathcal{L} _s\mathcal{L} _e}\mathcal{T} _e\mathcal{P} _e\mathbf{h}_{e}^{H}\mathbf{\Theta h}_r+n_e,
	\end{align}
	where $\mathbf{h}_{r}=[ h_{1}^{r},h_{2}^{r},...,h_{N}^{r} ] ^H\in \mathbb{C}^{N\times 1}$, $\mathbf{h}_{u}^{H}=[ h_{1}^{u},h_{2}^{u},...,h_{N}^{u} ] \in \mathbb{C}^{1\times N}$, $\mathbf{h}_{e}^{H}=[ h_{1}^{e},h_{2}^{e},...,h_{N}^{e} ] \in \mathbb{C}^{1\times N}$, $\mathbf{\Theta }=\mathrm{diag}\left( \mathbf{t} \right)$ and $\mathbf{t} =[ e^{j\upsilon _1},e^{j\upsilon _2},...,e^{j\upsilon _N} ] ^H\in \mathbb{C}^{N\times 1}$. Therefore, the SNR expressions (\ref{SNR_u}) and (\ref{SNR_e}) can also be re-expressed as
	\begin{align}
		\gamma _u=\overline{\gamma _{u}^{0}}\left| \mathbf{h}_{u}^{H}\mathbf{\Theta h}_{r} \right|^2, \\
		\gamma _e=\overline{\gamma _{e}^{0}}\left| \mathbf{h}_{e}^{H}\mathbf{\Theta h}_{r} \right|^2,
	\end{align}
	where $\overline{\gamma _{u}^{0}}=P\mathcal{L} _s\mathcal{L} _u\left( \mathcal{T} _u\mathcal{P} _u \right) ^2/\delta _{u}^{2}$ and $\overline{\gamma _{e}^{0}}=P\mathcal{L} _s\mathcal{L} _e\left( \mathcal{T} _e\mathcal{P} _e \right) ^2/\delta _{e}^{2}
	$, which can be regarded as constants for the variable $\mathbf{\Theta}$. To this end, the problem is formulated as
	\begin{align} 
		& \underset{\mathbf{\Theta }}{\max}\,\, \left[ \log _2\left( 1+\gamma _u \right) -\log _2\left( 1+\gamma _e \right) \right]^+ \label{p_1}
		\\
		& \,\mathrm{s.t.} \quad   \left| \left[ \mathbf{\Theta } \right] _{n,n} \right|=1, \,  n=1,2,...,N \tag{\ref{p_1}{a}}.
	\end{align}
	Next, we will discuss these two cases of E separately.
	
	\subsection{Optimization with Statistical CSI of E}
	With the statistical CSI of E, the problem (\ref{p_1}) can be transformed as follows:
	\begin{align} 
		& \underset{\mathbf{\Theta }}{\max } \,\, \frac{1+\overline{\gamma _{u}^{0}}\left| \mathbf{h}_{u}^{H}\mathbf{\Theta }\mathbf{h}_{r} \right|^2}{1+\overline{\gamma _{e}^{0}}N} \label{p_s_1}
		\\
		& \,\mathrm{s.t.}\quad \left| \left[ \mathbf{\Theta } \right] _{n,n} \right|=1,\, n=1,2,...,N\tag{\ref{p_s_1}{a}}. 
	\end{align}
	Apparently, the problem (\ref{p_s_1}) no longer has any connection with the CSI of E; therefore, it is wise to compensate for the cascaded channel phases of S-R and R-U links, that is, let $[ \mathbf{\Theta }^* ] _{n,n}=\exp ( -j\mathrm{arg}( [ \mathbf{h}_{r} ] _n+[ \mathbf{h}_{u} ] _n ) ) , \, n \in \{ 1,2,...,N \}$, where $\mathbf{\Theta}^*$ is the optimal solution to problem (\ref{p_s_1}). Then $\mathbf{t}^*$ is obtained by $[ \mathbf{t}^* ] _n=[ \mathbf{\Theta }^* ] _{n,n}, \, n \in \{1,2,...,N\}$. Limited by the finite adjustment precision of RIS, the final phase shifts can only be selected from a finite phase shift set $\mathbf{\Sigma }=\{ 0,\frac{2\pi}{2^b},...,\frac{( 2^b-1 ) 2\pi}{2^b} \}$. The optimal discreet phase shift of the $n$th element is denoted as $[\tilde{\mathbf{t}}^*]_n \in \mathbf{\Sigma}, \, n \in \left\{1,2,...,N\right\}$. Specifically, $[\tilde{\mathbf{t}}^*]_n$ is equal to the value which is closest to the optimal solution $\left[ \mathbf{t}^* \right] _n$ in the finite phase shift set. This operation of course should also think about the cyclic nature of phase shifts.
	
	\subsection{Optimization with Perfect CSI of E}
	With the perfect CSI of E, problem (\ref{p_1}) is transformed to 
	\begin{align}
		&\underset{\mathbf{\Theta }}{\max}\,\, \frac{1+\overline{\gamma _{u}^{0}}\left| \mathbf{h}_{u}^{H}\mathbf{\Theta h}_{r} \right|^2}{1+\overline{\gamma _{e}^{0}}\left| \mathbf{h}_{e}^{H}\mathbf{\Theta h}_{r} \right|^2}\label{p_i_1}
		\\
		&\,\mathrm{s.t.}\quad \left| \left[ \mathbf{\Theta } \right] _{n,n} \right|=1,\, n=1,2,...,N\tag{\ref{p_i_1}{a}},
	\end{align}	
	where the objective function and constraint are all non-convex. Applying the identical equation $\mathrm{diag}\left( \mathbf{a} \right) \mathbf{b}=\mathrm{diag}\left( \mathbf{b} \right) \mathbf{a}$ for two vectors $\mathbf{a}$ and $\mathbf{b}$, the above problem (\ref{p_i_1}) can be re-expressed as
	\begin{align}
		&\underset{\mathbf{t}}{\max}\,\,\frac{\mathbf{t}^H\varLambda _u\mathbf{t}}{\mathbf{t}^H\varLambda _e\mathbf{t}} \label{p_i_2}
		\\
		&\,\mathrm{s.t.} \quad \left| \left[ \mathbf{t} \right] _n \right|=1, \, n=1,2,...,N\tag{\ref{p_i_2}{a}},
	\end{align}
	where $\mathbf{t}$ is defined earlier at the beginning of the last subsection and
	\begin{align}
		\varLambda _u&=\frac{1}{N}+\overline{\gamma _{u}^{0}}\mathbf{\Omega }^H\mathbf{h}_{u}\mathbf{h}_{u}^{H}\mathbf{\Omega },\\
		\varLambda _e&=\frac{1}{N}+\overline{\gamma _{e}^{0}}\mathbf{\Omega }^H\mathbf{h}_{e}\mathbf{h}_{e}^{H}\mathbf{\Omega }, \\
		\mathbf{\Omega }&=\mathrm{diag}\left( \mathbf{h}_{r} \right).
	\end{align}
	It can be seen that the numerator and denominator of the objective function are both homogeneous quadratic, so that the semidefinite relaxation \cite{Luo2010Semidefinite} can be applied here. Specifically, let $\mathbf{\Psi }=\mathbf{tt}^H$. Therefore, $\mathbf{\Psi}$ is a semidefinite Hermitian matrix with rank $1$, and all diagonal elements of it are $1$ while the others equal $0$. In this case, problem (\ref{p_i_2}) can be further rewritten as
	\begin{align}
		&\max_{\mathbf{\Psi }} \,\, \frac{\mathrm{tr}\left( \varLambda _u\mathbf{\Psi } \right)}{\mathrm{tr}\left( \varLambda _e\mathbf{\Psi } \right)}\label{p_i_3}
		\\
		&\,\mathrm{s.t.}\quad    \left[\mathbf{\Psi }\right]_{n,n}=1, \, n=1,2,...,N,\tag{\ref{p_i_3}{a}}
		\\
		&\quad \quad \,\,\,\, \mathbf{\Psi }\succeq 0,\tag{\ref{p_i_3}{b}}\\
		&\quad \quad \,\,\,\, \mathrm{rank}\left( \mathbf{\Psi } \right) =1\tag{\ref{p_i_3}{c}}.
	\end{align}
	
	It is observed that the objective function in (\ref{p_i_3}) is the quotient of two linear functions. Therefore, we continue to make $\eta =1/\mathrm{tr}\left( \varLambda _e\mathbf{\Psi } \right)$. Meanwhile, to ensure that all constraints are convex, we temporarily relax (\ref{p_i_3}{c}) first. As a result, the problem (\ref{p_i_3}) can be re-expressed as
	\begin{align}
		&\max_{\eta ,\mathbf{\Phi }} \,\,\mathrm{tr}\left( \varLambda _u\mathbf{\Phi } \right) \label{p_i_4} 
		\\
		&\,\mathrm{s.t.} \quad  \eta \geqslant 0, \mathbf{\Phi }\succeq 0,\tag{\ref{p_i_4}{a}}
		\\
		& \quad \quad \,\,\, \mathrm{tr}\left( \varLambda _e\mathbf{\Phi } \right) =1,\tag{\ref{p_i_4}{b}}
		\\
		&\quad\quad\,\,      \left[ \mathbf{\Phi } \right] _{n,n}=\eta,\, n=1,2,...,N \tag{\ref{p_i_4}{c}},
	\end{align}
	where $\mathbf{\Phi }=\eta \mathbf{\Psi }$. The problem (\ref{p_i_4}) is a standard semidefinite programming problem, which can be solved by CVX. The optimal solutions to problem (\ref{p_i_4}) are denoted by $\eta^*$ and $\mathbf{\Phi}^*$, which satisfy $\mathbf{\Psi}^*=\mathbf{\Phi}^* / \eta^*$. However, due to the relaxation of (\ref{p_i_3}{c}), $\mathbf{\Psi}^*$ may not meet the rank-one constraint. Specifically, if $1<\mathrm{rank}\left( \mathbf{\Psi }^* \right) \leqslant N$, $\mathbf{\Psi }^*$ is the suboptimal solution to problem (\ref{p_i_3}), and $\mathbf{t }^*=\sqrt{e_{max}}\mathbf{q}_{max}$, where $e_{max}$ and $\mathbf{q}_{max}$ are the maximal eigenvalue of $\mathbf{\Psi}^*$ and the corresponding eigenvector to $e_{max}$, respectively; otherwise, the eigenvalue $e_{uni}$ of $\mathbf{\Psi }^*$ and corresponding eigenvector $\mathbf{q}_{uni}$ are unique. Therefore, $\mathbf{t}^*=\sqrt{e_{uni}}\mathbf{q}_{uni}$ is the optimal solution. The discretization of $\mathbf{t}^*$ is the same as that in the previous subsection to obtain $\tilde{\mathbf{t}}^*$.
	
	Our proposed optimization algorithms are summarized in Algorithm~\ref{alg1}. Since there is a specific optimal solution to problem (\ref{p_s_1}) and the objective function of problem (\ref{p_i_4}) is affine as well with a bounded variable, the convergence of  Algorithm~\ref{alg1} is guaranteed. Additionally, regarding to the  computational complexity, when only the statistical CSI of E is available, the phase shift of each element is derived in turn. Therefore, the problem (\ref{p_s_1}) can be solved with a complexity of $\mathcal{O} \left( N \right)$; on the other hand, the problem (\ref{p_i_4}) can be solved with a worst-case complexity of $\mathcal{O}\left(N^{4.5}\right)$ \cite{Luo2010Semidefinite}.
	
	\emph{Remark:} The optimization algorithms above are based on the assumption that there is only quantization error. Once the estimation error is also taken into account along with the quantization error, two independent phase errors subject to the Von Mises distribution should be imposed to the quantized optimal phase shifts before applied to S-R-U and S-R-E links, respectively. The related results will be shown in Section~V.
	
	\begin{algorithm}[!t] 
		\renewcommand{\algorithmicrequire}{\textbf{Input:}} 
		\renewcommand{\algorithmicensure}{\textbf{Output:}} 
		\caption{Phase Shifts Optimization Algorithm to Maximize the SR}
		\label{alg1}
		\begin{algorithmic}[1] 
			\REQUIRE
			Channels $\mathbf{h}_{r}$, $\mathbf{h}_{u}$, $\mathbf{h}_{e}$; Constants $\overline{\gamma_{u}^0}$, $\overline{\gamma_{e}^0}$
			\ENSURE
			$\tilde{\mathbf{t}}^*$ 
			\\
			\IF{only the statistical CSI of E is available}
			\STATE $\left[ \mathbf{\Theta }^* \right] _{n,n}=\exp ( -j\mathrm{arg}( [ \mathbf{h}_{r} ] _n+\left[ \mathbf{h}_{u} \right] _n ) ) ,\, n=1,2,...,N $
			\STATE $[ \mathbf{t}^* ] _n=[ \mathbf{\Theta }^* ] _{n,n}, \, n=1,2,...,N $
			\ELSE
			\STATE solve problem (\ref{p_i_4}) to find $\eta^*$ and $\mathbf{\Phi}^*$
			\STATE $\mathbf{\Psi}^*=\mathbf{\Phi}^* / \eta^*$
			\STATE execute eigenvalue decomposition of $\mathbf{\Psi}^*$ to obtain its eigenvalues $\{ e _i \} ,\, 1\leqslant i\leqslant N$ and corresponding eigenvectors $\{ \mathbf{q}_i \} ,\, 1\leqslant i\leqslant N$			
			\IF{$1<\mathrm{rank}( \mathbf{\Psi }^* ) \leqslant N$}
			\STATE $\mathbf{t }^*=\sqrt{e_{max}}\mathbf{q}_{max}$
			\ELSE
			\STATE $\mathbf{t}^*=\sqrt{e_{uni}}\mathbf{q}_{uni}$
			\ENDIF
			\ENDIF
			\STATE discretize $\mathbf{t}^*$ based on the finite phase shift set $\mathbf{\Sigma }=\{ 0,\frac{2\pi}{2^b},...,\frac{( 2^b-1 ) 2\pi}{2^b} \}$ to yield $\tilde{\mathbf{t}}^*$
			\RETURN $\tilde{\mathbf{t}}^*$
		\end{algorithmic}
	\end{algorithm}

	\section{Simulation Results}
	
	To validate the theoretical analysis and evaluate the performance of the proposed algorithms, in this section, numerical results are presented under simulated channels. All simulations are run over $10000$ times.
	
	Unless otherwise specified, the parameters used in the simulations are set as follows: the operating wavelength $\lambda=500$~$\mathrm{\mu}$m (the corresponding frequency is $0.6$~THz); the distance between two communicating nodes $d_{sr}=150$~km, $d_{ru}=19$~km, $d_{re}=20$~km; the beam waist at U and E $w _u=w _e=w$; the antenna gains $G_s = 69$~dBi, $G_u = G_e = 55$~dBi (inferring the equal radius of the effective area of receive antennas on U and E $l_u = l_e = l$); the noise power at receivers $\delta_{u}^2=\delta_{e}^2=\delta^2$. Besides, for brevity, the single quantization error case and two-error case are represented by P1 and P2 in the legends of all figures, respectively.
	
	\begin{figure}[!t]
		\centering
		\includegraphics[width=3.0in]{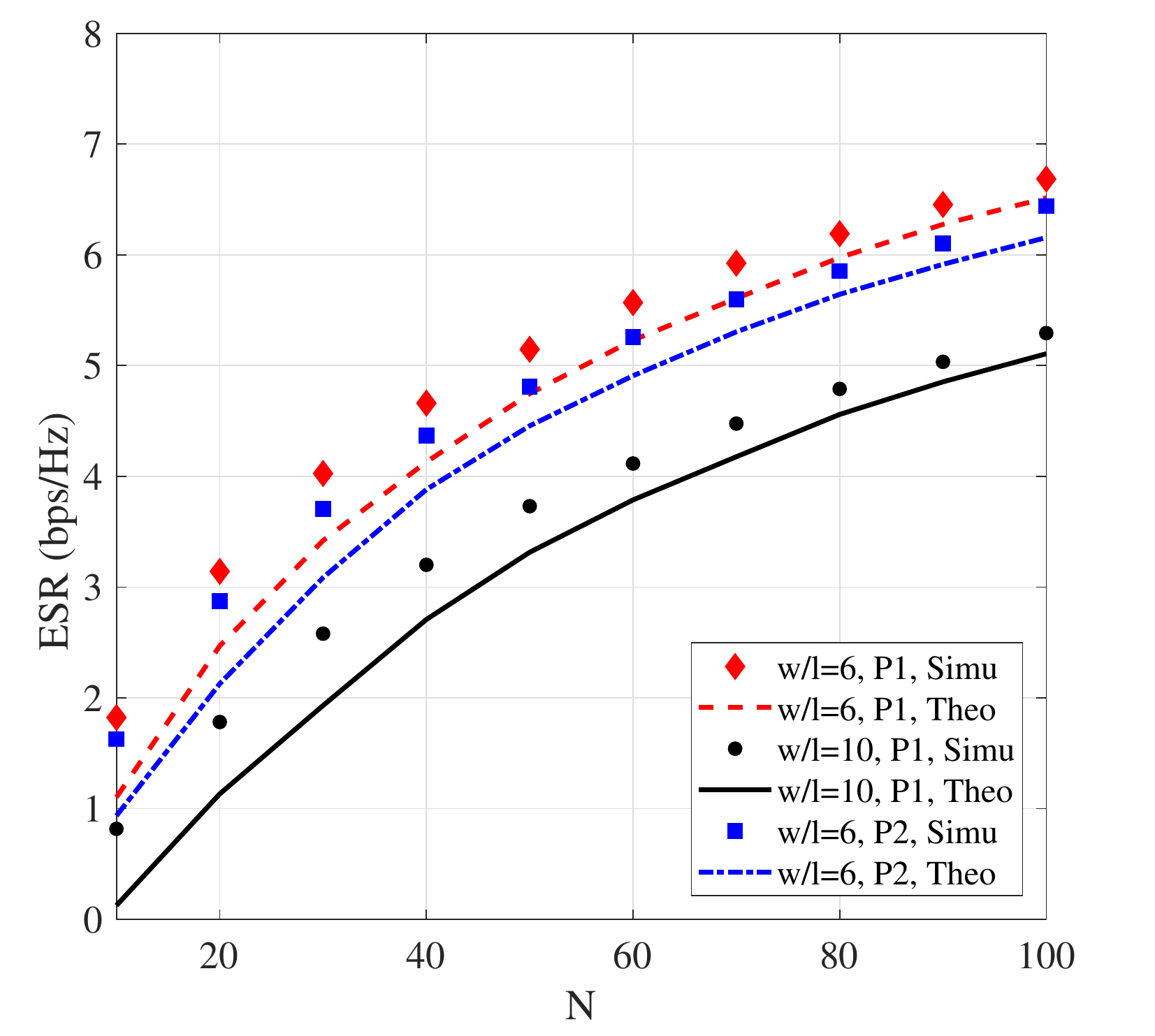}
		\caption{Simulation and theoretical results for ESR versus the number of element $N$ with $b=1$, $\kappa=5$, $\sigma_j^{u}=0.1$, $\sigma_j^e=0.2$, $C_n^2=10^{-13}$, $P/\delta^2=260$~dB.}
		\label{ESR_sim_theo}
	\end{figure}
	
	Fig.~\ref{ESR_sim_theo} shows a comparison of Monte Carlo simulations and theoretical results for the ESR versus the number of reflective elements $N$. It can be seen that as $N$ increases, the gap between the theoretical ESR and the simulation results decreases, no matter there is P1 or P2. In addition, it is evident that the growth of $N$ brings an increase in ESR. For instance, under the case of P1 with $w/l=6$, the ESR equals about $1.9$~bps/Hz when $N=10$, while it increases to around $6.6$~bps/Hz when $N=100$, indicating the diversity gain brought by RIS. In contrast, a larger ratio of $w$ to $l$ leads to a smaller ESR, which is because an excessively large $w/l$ reduces the energy per unit area of the wave reaching the receivers.
	
	\begin{figure}[!t]
		\centering
		\includegraphics[width=3.0in]{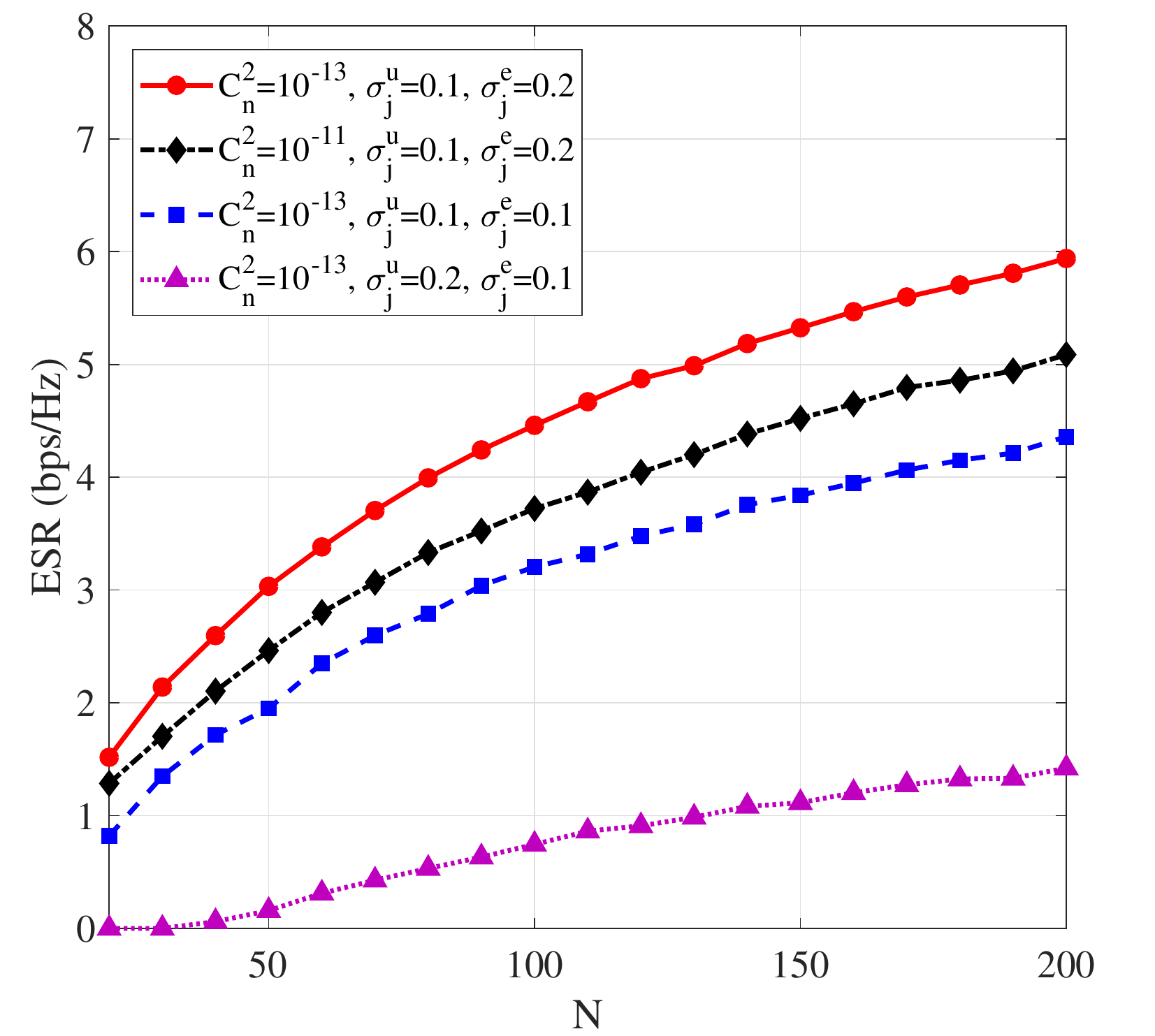}
		\caption{ESR versus the number of element $N$ under two-error case with $w/l=6$, $b=1$, $\kappa=1$, $P/\delta^2=260$~dB.}
		\label{ESR_paras}
	\end{figure}
	
	\begin{figure}[!t]
		\begin{subfigure}{.23\textwidth}
			\centering
			\includegraphics[width=1.7in]{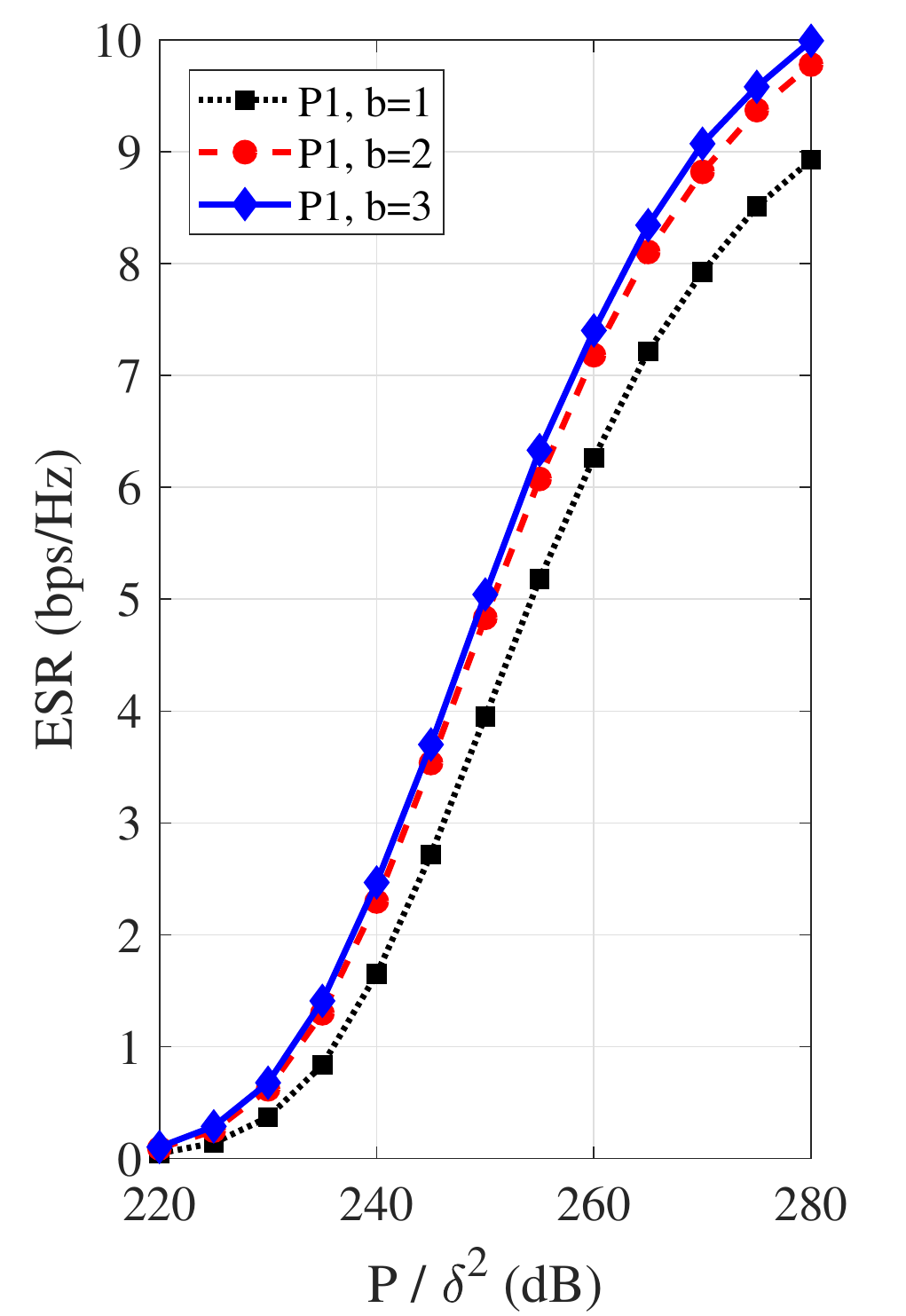}
			\caption{Single quantization error.}
			\label{ESR_k_b.a}
		\end{subfigure}
		\begin{subfigure}{.23\textwidth}
			\centering
			\includegraphics[width=1.7in]{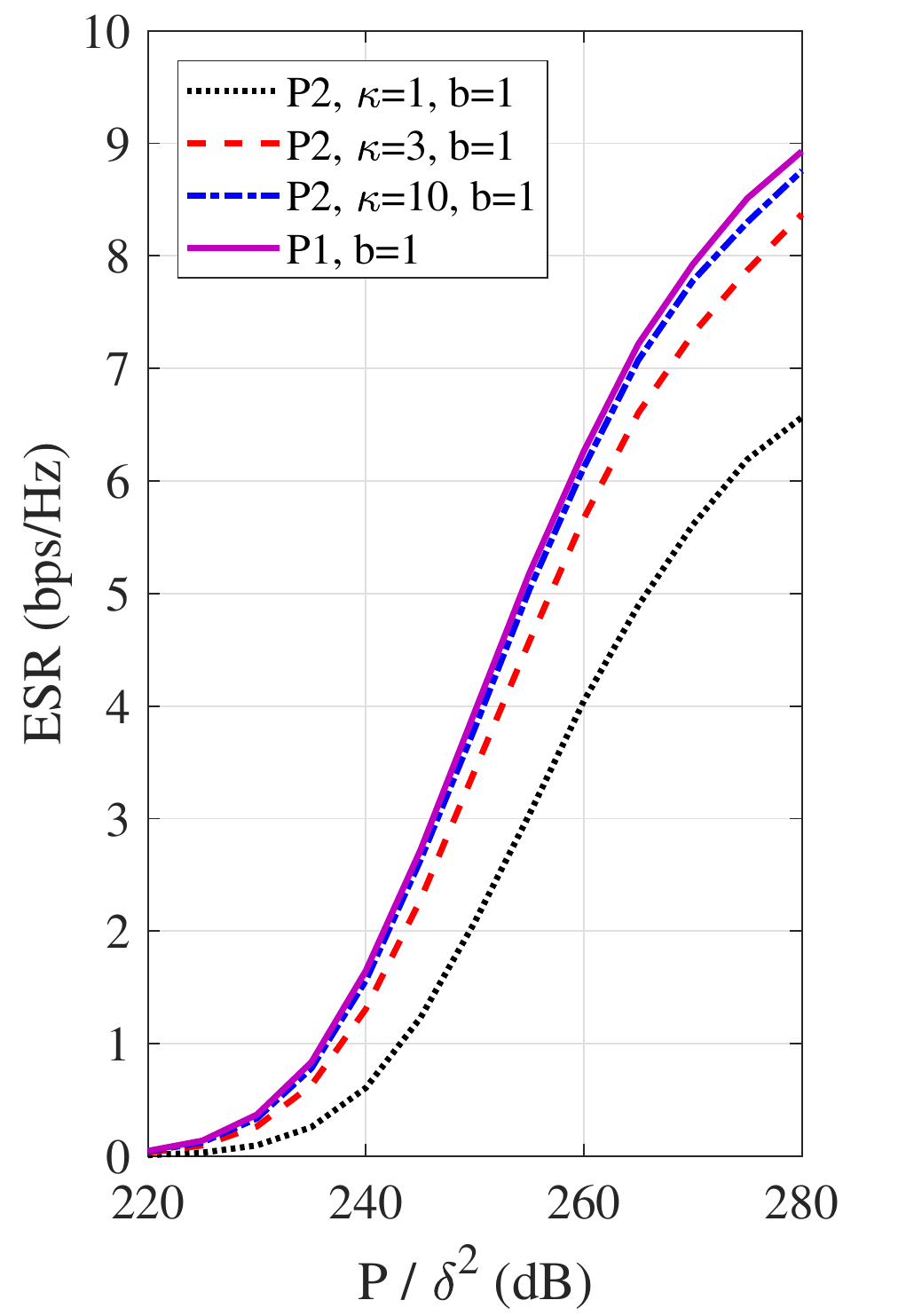}
			\caption{Two-error case with $b=1$.}
			\label{ESR_k_b.b}
		\end{subfigure}
		\caption{ESR versus transmit SNR with $N=80$, $w/l=6$, $\sigma_j^u=0.1$, $\sigma_j^{e}=0.2$, $C_n^2=10^{-13}$.}
		\label{ESR_k_b}
	\end{figure}
	
	As depicted in Fig.~\ref{ESR_paras}, the index of refraction structure parameter $C_n^2$ going up means more intense atmosphere turbulence, and as a result the ESR eventually drops. Moreover, it is intuitive that the ESR can be generally guaranteed to be greater than $0$ when the jitter standard variance of U is smaller than that of E, as long as the other parameters are exactly the same. Whereas Fig.~\ref{ESR_paras} shows that it is still practicable to remain the ESR positive when $\sigma_j^u=\sigma_j^e$ and even when $\sigma_j^u > \sigma_j^e$ under the worst phase error case with $b=1$ and $\kappa=1$. Specifically, when $\sigma_j^u=0.2$ and $\sigma_j^e=0.1$, the ESR increases from near $0$~bps/Hz to almost $1.4$~bps/Hz as $N$ increases from $20$ to $200$. Therefore, by increasing the number of element, the system security can be ensured even in a hostile environment.
	
	Additionally, we investigate the relationships between the quantization bit $b$ of uniform distribution, the concentration parameter $\kappa$ of Von Mises distribution, and the ESR in Fig.~\ref{ESR_k_b}. Specifically, the ESR improves as $\kappa$ or $b$ increases, as a result of a decreasing phase error. However, it is worth noting that the larger the values of  $b$ and $\kappa$, the weaker the effect of increasing them on system security. The main reason is that when there is a high-precision RIS or the knowledge of CSI is abundant ($b$ or $\kappa$ is big), it does not make much sense to continue to enhance them. Furthermore, Fig.~\ref{ESR_k_b}(\subref{ESR_k_b.b}) shows that the curves under P2 with $\kappa=10$ and $b = 1$ is very close to that under P1 with $b=1$, and it can be inferred from their relationship that these two curves will definitely coincide as long as $\kappa$ becomes large enough. Moreover, the coordinates in Fig.~(\ref{ESR_k_b}) imply a positive correlation between the ESR and transmit SNR.
	
	\begin{figure}[!t]
		\begin{subfigure}{.23\textwidth}
			\centering
			\includegraphics[width=1.7in]{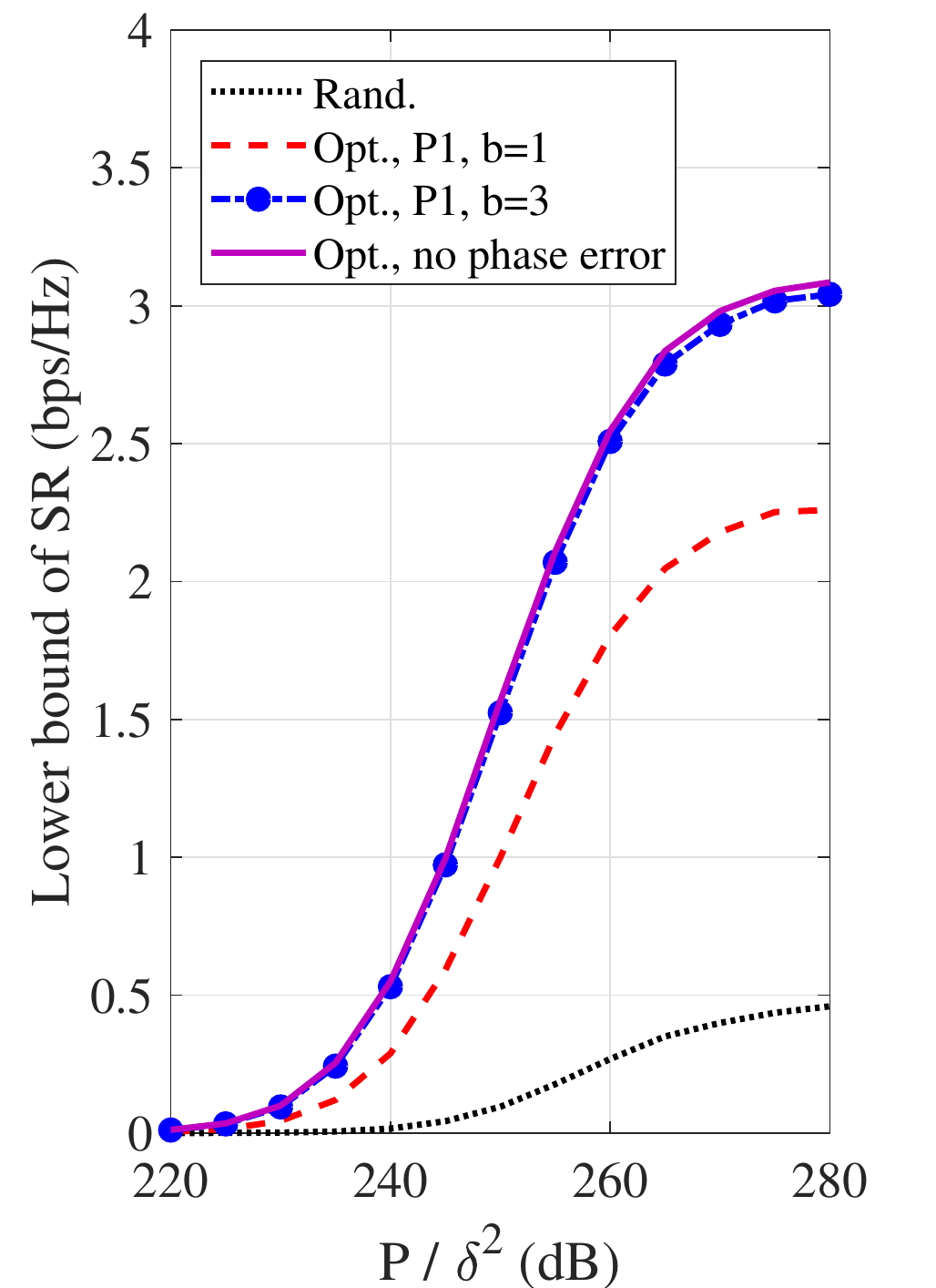}
			\caption{Single quantization error.}
			\label{opt_sr.a}
		\end{subfigure}
		\begin{subfigure}{.23\textwidth}
			\centering
			\includegraphics[width=1.7in]{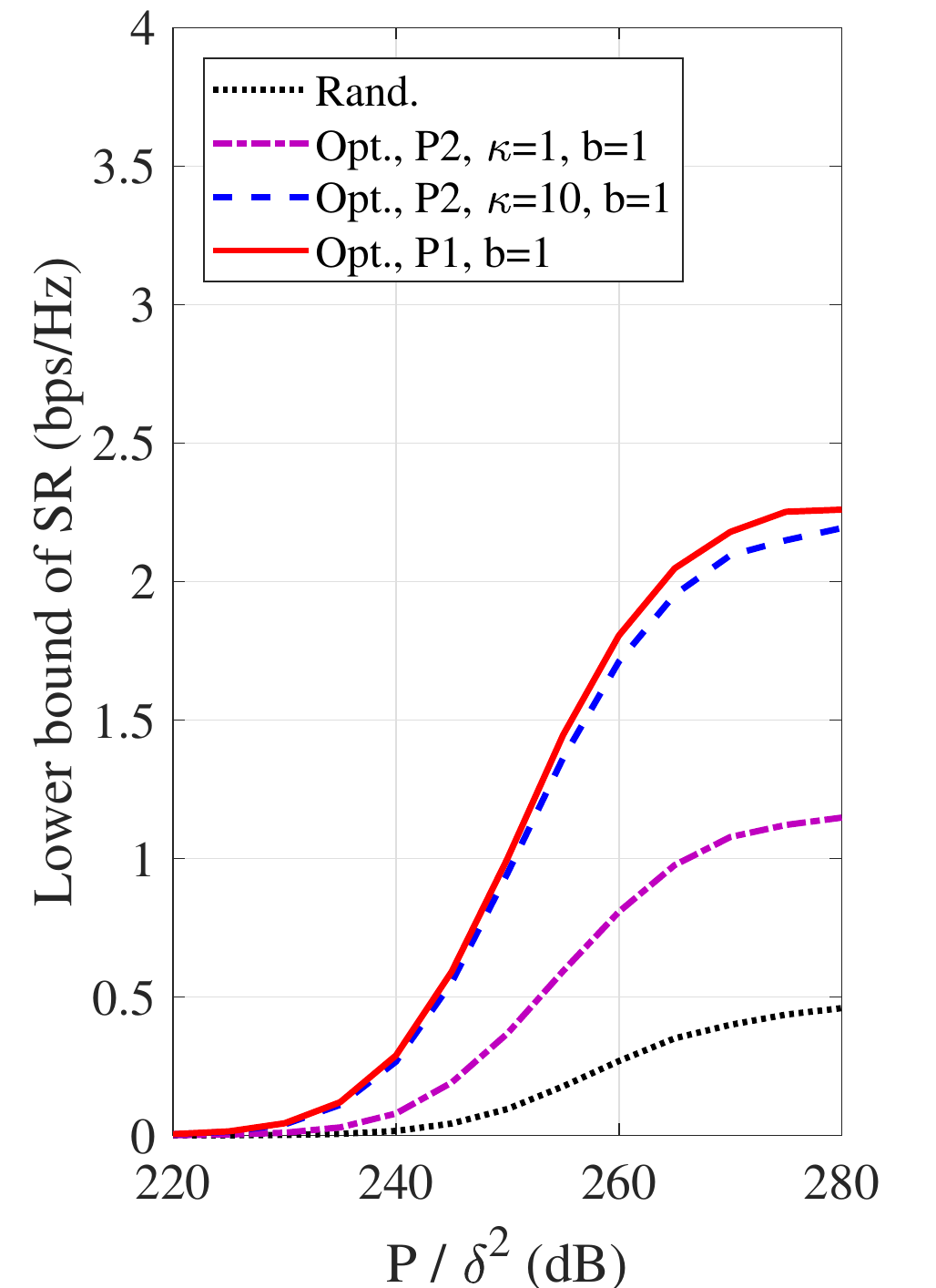}
			\caption{Two-error case with $b=1$.}
			\label{opt_sr.b}
		\end{subfigure}
		\newline	
		\begin{subfigure}{.23\textwidth}
			\centering
			\includegraphics[width=1.7in]{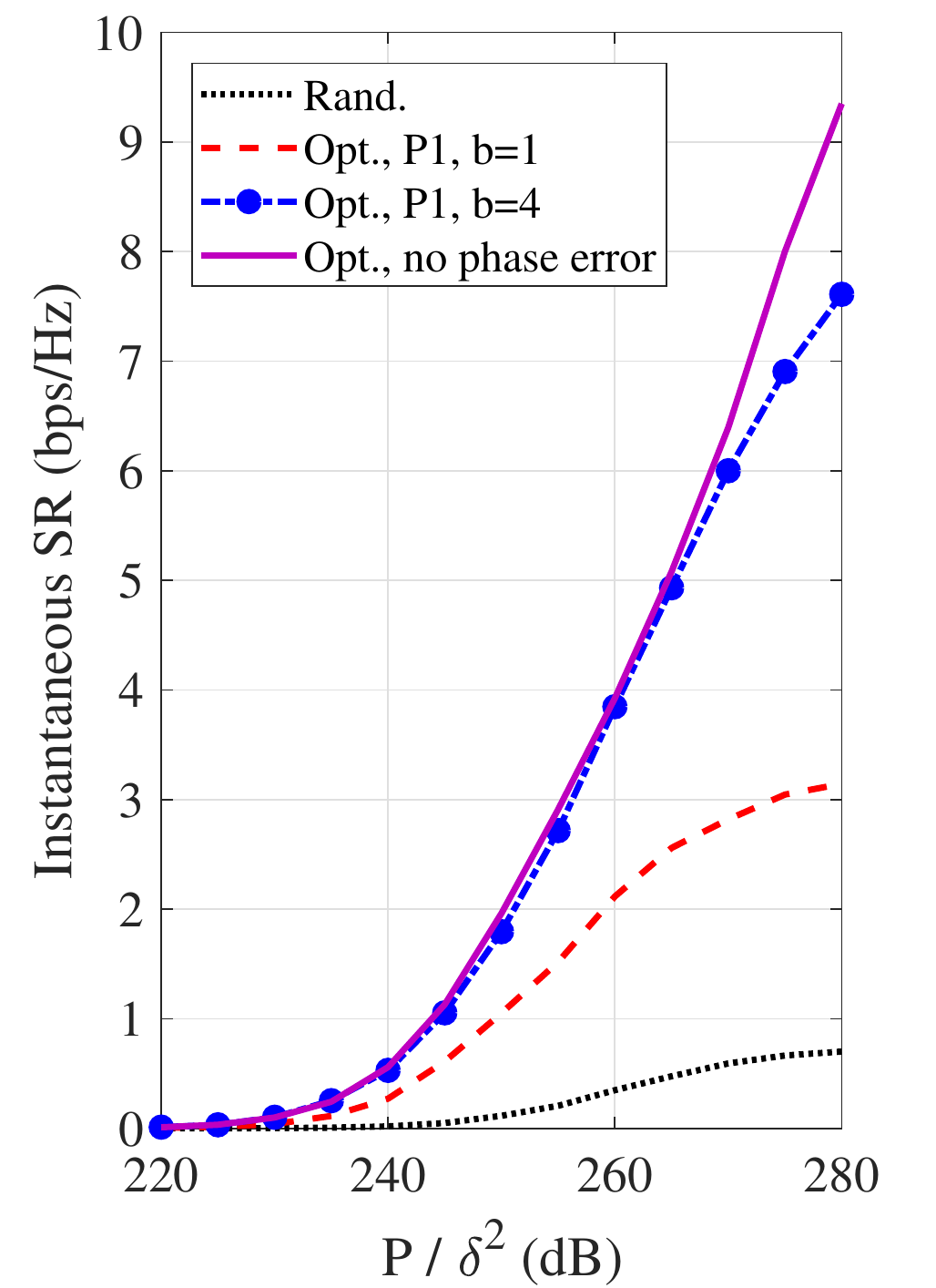}
			\caption{Single quantization error.}
			\label{opt_sr.c}
		\end{subfigure}
		\begin{subfigure}{.23\textwidth}
			\centering
			\includegraphics[width=1.7in]{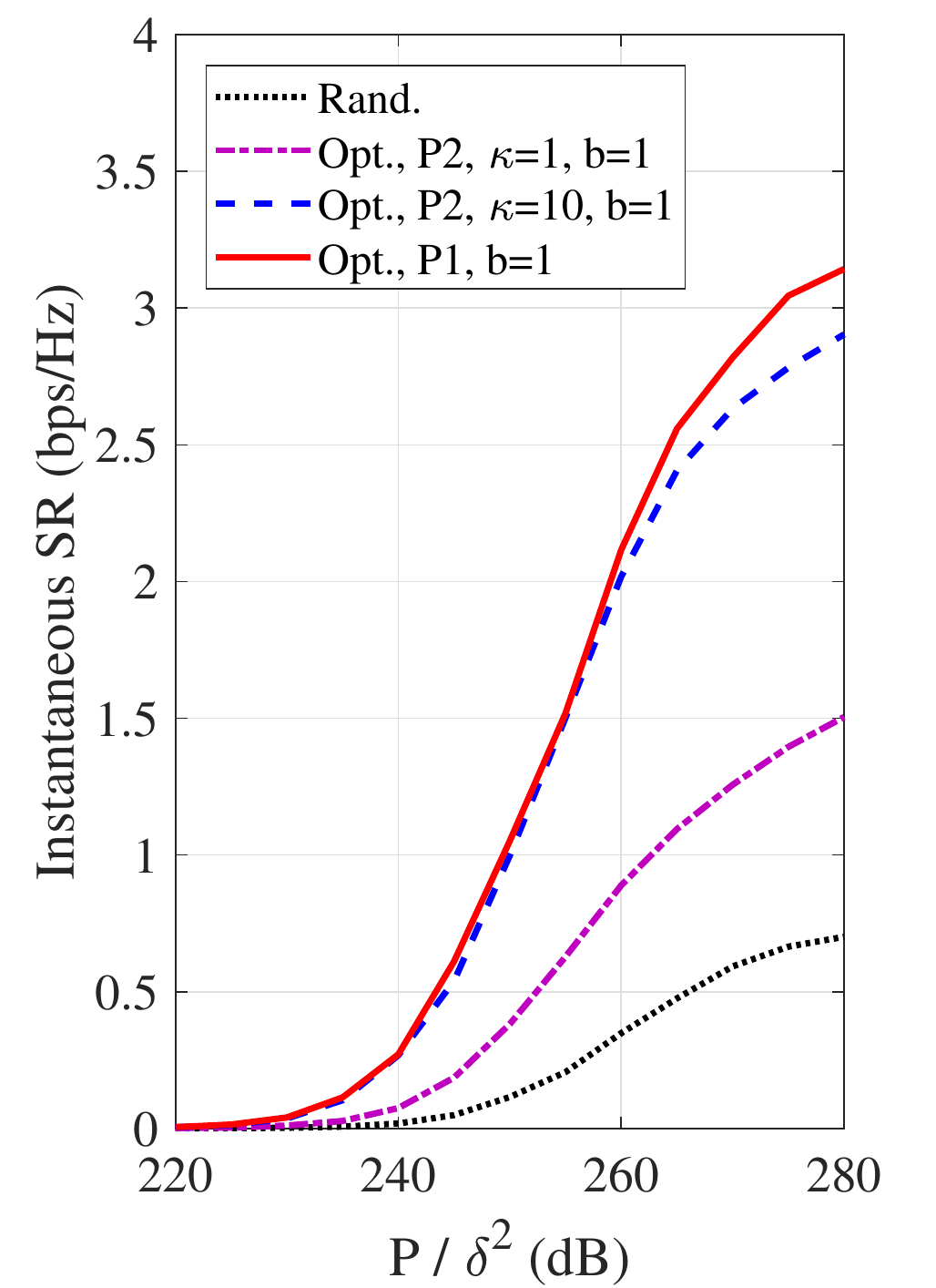}
			\caption{Two-error case with $b=1$.}
			\label{opt_sr.d}
		\end{subfigure}
		\caption{The optimized lower bound of SR and instantaneous SR versus transmit SNR with $N=40$, $w/l=6$, $\sigma_j^u=0.2$, $\sigma_j^e=0.1$, and $C_n^2=10^{-13}$.}
		\label{opt_sr}
	\end{figure}
	
	Fig.~\ref{opt_sr} presents the optimized lower bound of SR and instantaneous SR, respectively. The abbreviations Rand. and Opt. in the legends represent random and optimized phase shifts, respectively. As seen from this figure, the system is in a serious secrecy risk when phase shifts of RIS are randomly selected. However, both the lower bound of SR and instantaneous SR are greatly improved with the help of our proposed optimization algorithms. Specifically, the system achieves the best secrecy performance when the phase shifts of all elements are optimized without phase error, where the lower bound of SR and instantaneous SR increase by around $2.25$~bps/Hz and $3.5$~bps/Hz respectively when the transmit SNR is $260$~dB compared to the random phase shift case. Moreover, when the number of quantization bits $b$ reaches $3$, the optimized lower bound of SR obtained with a single quantization error is almost equal to the best result, and all of them tend to be stable as the transmit SNR increases. However, it needs a higher quantization precision to get closer to the best result for the instantaneous SR. Similarly, under P2 with $b=1$, compared to the lower bound of SR, the instantaneous SR needs a larger $\kappa$ to reach the upper bound of it, which means more accurate CSI. Finally, Fig.~\ref{opt_sr} also reveals that the system can be maintained at a high level of security (both the lower bound of SR and instantaneous SR are far greater than $0$) by taking advantage of the proposed optimization algorithms even when $\sigma_j^u > \sigma_j^e$.
	
	\section{Conclusion}
	In this paper, we investigated a novel secure THz-empowered RANTN, where a LEO satellite indirectly send confidential information to a UAV in the troposphere, bridging by a RIS-mounted HAP situated in the stratosphere, while there is another untrustworthy UAV receiver near the trustworthy one. On account of the atmospheric environment, the sharp beam of THz wave, the relative motion of two communicating nodes, the finite adjustment precision of RIS and the imperfect channel estimation, the fading effects resulting from atmosphere turbulence, pointing error and phase error were all taken into account simultaneously. Then the approximate ESR expression was derived. Besides, the phase shifts of RIS were optimized to enhance the security of the system with statistical and perfect CSI of the untrustworthy UAV, respectively. Finally, simulation results verified the effectiveness of our theoretical analysis and optimization algorithms.

	
	\bibliographystyle{IEEEtran}
	\bibliography{TRANTN-ref}

\begin{thebibliography}{10}
\providecommand{\url}[1]{#1}
\csname url@samestyle\endcsname
\providecommand{\newblock}{\relax}
\providecommand{\bibinfo}[2]{#2}
\providecommand{\BIBentrySTDinterwordspacing}{\spaceskip=0pt\relax}
\providecommand{\BIBentryALTinterwordstretchfactor}{4}
\providecommand{\BIBentryALTinterwordspacing}{\spaceskip=\fontdimen2\font plus
\BIBentryALTinterwordstretchfactor\fontdimen3\font minus
  \fontdimen4\font\relax}
\providecommand{\BIBforeignlanguage}[2]{{%
\expandafter\ifx\csname l@#1\endcsname\relax
\typeout{** WARNING: IEEEtran.bst: No hyphenation pattern has been}%
\typeout{** loaded for the language `#1'. Using the pattern for}%
\typeout{** the default language instead.}%
\else
\language=\csname l@#1\endcsname
\fi
#2}}
\providecommand{\BIBdecl}{\relax}
\BIBdecl

\bibitem{Giordani2021Non}
M.~Giordani and M.~Zorzi, ``Non-terrestrial networks in the 6{G} era:
  Challenges and opportunities,'' \emph{IEEE Netw.}, vol.~35, no.~2, pp.
  244--251, Mar./Apr. 2021.

\bibitem{Leyva2021Inter}
I.~Leyva-Mayorga, B.~Soret, and P.~Popovski, ``Inter-plane inter-satellite
  connectivity in dense {LEO} constellations,'' \emph{IEEE Trans. Wireless
  Commun.}, vol.~20, no.~6, pp. 3430--3443, Jun. 2021.

\bibitem{Le2022Throughput}
H.~D. Le, P.~V. Trinh, T.~V. Pham, D.~R. Kolev, A.~Carrasco-Casado,
  T.~Kubo-Oka, M.~Toyoshima, and A.~T. Pham, ``Throughput analysis for {TCP}
  over the {FSO}-based satellite-assisted {I}nternet of vehicles,'' \emph{IEEE
  Trans. Veh. Technol.}, vol.~71, no.~2, pp. 1875--1890, Feb. 2022.

\bibitem{Jia2021Joint}
Z.~Jia, M.~Sheng, J.~Li, D.~Zhou, and Z.~Han, ``Joint {HAP} access and {LEO}
  satellite backhaul in 6{G}: Matching game-based approaches,'' \emph{IEEE J.
  Sel. Areas Commun.}, vol.~39, no.~4, pp. 1147--1159, Aug. 2021.

\bibitem{Saeed2021Point}
N.~Saeed, H.~Almorad, H.~Dahrouj, T.~Y. Al-Naffouri, J.~S. Shamma, and M.-S.
  Alouini, ``Point-to-point communication in integrated satellite-aerial 6{G}
  networks: State-of-the-art and future challenges,'' \emph{IEEE Open J.
  Commun. Society}, Jun. 2021.

\bibitem{Kaushal2017Optical}
H.~Kaushal and G.~Kaddoum, ``Optical communication in space: Challenges and
  mitigation techniques,'' \emph{IEEE Commun. Surveys Tuts.}, vol.~19, no.~1,
  pp. 57--96, Aug. 2017.

\bibitem{Kokkoniemi2021Channel}
J.~Kokkoniemi, J.~M. Jornet, V.~Petrov, Y.~Koucheryavy, and M.~Juntti,
  ``Channel modeling and performance analysis of airplane-satellite terahertz
  band communications,'' \emph{IEEE Trans. Veh. Technol.}, vol.~70, no.~3, pp.
  2047--2061, Mar. 2021.

\bibitem{Ntontin2017Toward}
K.~Ntontin and C.~Verikoukis, ``Toward the performance enhancement of microwave
  cellular networks through {THz} links,'' \emph{IEEE Trans. Veh. Technol.},
  vol.~66, no.~7, pp. 5635--5646, Jul. 2017.

\bibitem{Bhardwaj2021Performance}
P.~Bhardwaj and S.~M. Zafaruddin, ``Performance of dual-hop relaying for
  {THz-RF} wireless link over asymmetrical $\alpha$-$\mu$ fading,'' \emph{IEEE
  Trans. Veh. Technol.}, vol.~70, no.~10, pp. 10\,031--10\,047, Oct. 2021.

\bibitem{Andrews1999Theory}
L.~C. Andrews, R.~L. Phillips, C.~Y. Hopen, and M.~A. Al-Habash, ``Theory of
  optical scintillation,'' \emph{J. Opt. Soc. Am. A}, vol.~16, no.~6, pp.
  1417--1429, Jun. 1999.

\bibitem{Ammar2001Mathematical}
A.~Al-Habash, L.~C. Andrews, and R.~L. Phillips, ``{Mathematical model for the
  irradiance probability density function of a laser beam propagating through
  turbulent media},'' \emph{Opt. Eng.}, vol.~40, no.~8, pp. 1554 -- 1562, Aug.
  2001.

\bibitem{Wu2020towards}
Q.~Wu and R.~Zhang, ``Towards smart and reconfigurable environment: Intelligent
  reflecting surface aided wireless network,'' \emph{IEEE Commun. Mag.},
  vol.~58, no.~1, pp. 106--112, Jan. 2020.

\bibitem{Ye2021Non}
J.~{Ye}, J.~{Qiao}, A.~{Kammoun}, and M.-S. {Alouini}, ``Non-terrestrial
  communications assisted by reconfigurable intelligent surfaces,'' \emph{arXiv
  e-prints, arXiv:2109.00876}, Sept. 2021.

\bibitem{Jia2020Ergodic}
H.~Jia, J.~Zhong, M.~N. Janardhanan, and G.~Chen, ``Ergodic capacity analysis
  for {FSO} communications with {UAV}-equipped {IRS} in the presence of
  pointing error,'' in \emph{Proc. IEEE Int. Conf. Commun. Technol. (ICCT)},
  Oct. 2020, pp. 949--954.

\bibitem{Tekbiyik2020Reconfigurable}
K.~{Tekb{\i}y{\i}k}, G.~{Karabulut Kurt}, A.~{R{\i}za Ekti},
  A.~{G{\"o}r{\c{c}}in}, and H.~{Yanikomeroglu}, ``{Reconfigurable intelligent
  surfaces empowered THz communication in LEO satellite networks},''
  \emph{arXiv e-prints, arXiv: 2007.04281}, Jul. 2020.

\bibitem{Li2019Secret}
K.~Li, L.~Lu, W.~Ni, E.~Tovar, and M.~Guizani, ``Secret key agreement for data
  dissemination in vehicular platoons,'' \emph{IEEE Trans. Veh. Technol.},
  vol.~68, no.~9, pp. 9060--9073, Sept. 2019.

\bibitem{Kawai2021QoS}
Y.~Kawai and S.~Sugiura, ``{QoS}-constrained optimization of intelligent
  reflecting surface aided secure energy-efficient transmission,'' \emph{IEEE
  Trans. Veh. Technol.}, vol.~70, no.~5, pp. 5137--5142, May 2021.

\bibitem{Dong2020Enhancing}
L.~Dong and H.-M. Wang, ``Enhancing secure {MIMO} transmission via intelligent
  reflecting surface,'' \emph{IEEE Trans. Wireless Commun.}, vol.~19, no.~11,
  pp. 7543--7556, Nov. 2020.

\bibitem{Wang2022Beamforming}
W.~Wang, X.~Liu, J.~Tang, N.~Zhao, Y.~Chen, Z.~Ding, and X.~Wang, ``Beamforming
  and jamming optimization for {IRS}-aided secure {NOMA} networks,'' \emph{IEEE
  Trans. Wireless Commun.}, vol.~21, no.~3, pp. 1557--1569, Mar. 2022.

\bibitem{Guan2020Intelligent}
X.~Guan, Q.~Wu, and R.~Zhang, ``Intelligent reflecting surface assisted secrecy
  communication: Is artificial noise helpful or not?'' \emph{IEEE Wireless
  Commun. Lett.}, vol.~9, no.~6, pp. 778--782, Jun. 2020.

\bibitem{Badiu2020Communication}
M.-A. Badiu and J.~P. Coon, ``Communication through a large reflecting surface
  with phase errors,'' \emph{IEEE Wireless Commun. Lett.}, vol.~9, no.~2, pp.
  184--188, Feb. 2020.

\bibitem{Trigui2021Secrecy}
I.~Trigui, W.~Ajib, and W.-P. Zhu, ``Secrecy outage probability and average
  rate of {RIS}-aided communications using quantized phases,'' \emph{IEEE
  Commun. Lett.}, vol.~25, no.~6, pp. 1820--1824, Jun. 2021.

\bibitem{Vega2021Physical}
J.~D. Vega~Sánchez, P.~Ramírez-Espinosa, and F.~J. López-Martínez,
  ``Physical layer security of large reflecting surface aided communications
  with phase errors,'' \emph{IEEE Wireless Commun. Lett.}, vol.~10, no.~2, pp.
  325--329, Feb. 2021.

\bibitem{Xu2021Ergodic}
P.~Xu, G.~Chen, G.~Pan, and M.~D. Renzo, ``Ergodic secrecy rate of
  {RIS}-assisted communication systems in the presence of discrete phase shifts
  and multiple eavesdroppers,'' \emph{IEEE Wireless Commun. Lett.}, vol.~10,
  no.~3, pp. 629--633, Mar. 2021.

\bibitem{Fang2021Joint}
S.~Fang, G.~Chen, and Y.~Li, ``Joint optimization for secure intelligent
  reflecting surface assisted {UAV} networks,'' \emph{IEEE Wireless Commun.
  Lett.}, vol.~10, no.~2, pp. 276--280, Feb. 2021.

\bibitem{Li2021Robust}
S.~Li, B.~Duo, M.~D. Renzo, M.~Tao, and X.~Yuan, ``Robust secure {UAV}
  communications with the aid of reconfigurable intelligent surfaces,''
  \emph{IEEE Trans. Wireless Commun.}, vol.~20, no.~10, pp. 6402--6417, Oct.
  2021.

\bibitem{Guo2021Learning}
X.~Guo, Y.~Chen, and Y.~Wang, ``Learning-based robust and secure transmission
  for reconfigurable intelligent surface aided millimeter wave {UAV}
  communications,'' \emph{IEEE Wireless Commun. Lett.}, vol.~10, no.~8, pp.
  1795--1799, Aug. 2021.

\bibitem{Li2020A}
Y.~Li, N.~Deng, and W.~Zhou, ``A hierarchical approach to resource allocation
  in extensible multi-layer {LEO-MSS},'' \emph{IEEE Access}, vol.~8, pp.
  18\,522--18\,537, Jan. 2020.

\bibitem{Karabulut2021A}
G.~Karabulut~Kurt, M.~G. Khoshkholgh, S.~Alfattani, A.~Ibrahim, T.~S.~J.
  Darwish, M.~S. Alam, H.~Yanikomeroglu, and A.~Yongacoglu, ``A vision and
  framework for the high altitude platform station ({HAP}s) networks of the
  future,'' \emph{IEEE Commun. Surveys Tuts.}, vol.~23, no.~2, pp. 729--779,
  Mar. 2021.

\bibitem{Goldsmith2005Wireless}
A.~Goldsmith, \emph{Wireless Communications}.\hskip 1em plus 0.5em minus
  0.4em\relax New York, USA: Cambridge University Press, 2005.

\bibitem{Le2021Performance}
M.~Le-Tran, T.-H. Vu, and S.~Kim, ``Performance analysis of optical backhauled
  cooperative {NOMA} visible light communication,'' \emph{IEEE Trans. Veh.
  Technol.}, vol.~70, no.~12, pp. 12\,932--12\,945, Dec. 2021.

\bibitem{Najafi2021Intelligent}
M.~Najafi, B.~Schmauss, and R.~Schober, ``Intelligent reflecting surfaces for
  free space optical communication systems,'' \emph{IEEE Trans. Commun.},
  vol.~69, no.~9, pp. 6134--6151, Sept. 2021.

\bibitem{Ning2021Terahertz}
B.~Ning, Z.~Chen, W.~Chen, Y.~Du, and J.~Fang, ``Terahertz multi-user massive
  {MIMO} with intelligent reflecting surface: Beam training and hybrid
  beamforming,'' \emph{IEEE Trans. Veh. Technol.}, vol.~70, no.~2, pp.
  1376--1393, Feb. 2021.

\bibitem{Priebe2013Ultra}
S.~Priebe, M.~Kannicht, M.~Jacob, and T.~Kürner, ``Ultra broadband indoor
  channel measurements and calibrated ray tracing propagation modeling at {THz}
  frequencies,'' \emph{J. Commun. Net.}, vol.~15, no.~6, pp. 547--558, Dec.
  2013.

\bibitem{Wang2021Outage}
T.~Wang, M.-A. Badiu, G.~Chen, and J.~P. Coon, ``Outage probability analysis of
  {RIS}-assisted wireless networks with {Von Mises} phase errors,'' \emph{IEEE
  Wireless Commun. Lett.}, Early Access, 2021.

\bibitem{Sandalidis2016A}
H.~G. Sandalidis, N.~D. Chatzidiamantis, and G.~K. Karagiannidis, ``A tractable
  model for turbulence- and misalignment-induced fading in optical wireless
  systems,'' \emph{IEEE Commun. Lett.}, vol.~20, no.~9, pp. 1904--1907, Sept.
  2016.

\bibitem{Milton1972Handbook}
M.~Abramowitz and I.~A. Stegun, \emph{Handbook of Mathematical Functions: With
  Formulas, Graphs, and Mathematical Tables, 9th ed.}\hskip 1em plus 0.5em
  minus 0.4em\relax New York, USA: Dover, 1972.

\bibitem{Gappmair2017Subcarrier}
W.~Gappmair and H.~E. Nistazakis, ``Subcarrier {PSK} performance in terrestrial
  {FSO} links impaired by {Gamma-Gamma} fading, pointing errors, and phase
  noise,'' \emph{J. Lightwave Technol.}, vol.~35, no.~9, pp. 1624--1632, May
  2017.

\bibitem{Luo2010Semidefinite}
Z.~Luo, W.~K. Ma, A.~M. cho So, Y.~Ye, and S.~Zhang, ``Semidefinite relaxation
  of quadratic optimization problems,'' \emph{IEEE Signal Process. Mag.},
  vol.~27, no.~3, pp. 20--34, May 2010.

\end{thebibliography}
	
\end{document}